\documentclass[a4paper,11pt]{article}
\usepackage{amsmath}
\usepackage{graphicx}
\usepackage{amsbsy}
\usepackage{cite}
\usepackage{slashed}
\usepackage{amsfonts}
\usepackage{amssymb}

\begin{document}

\title{Could a light Higgs boson illuminate the dark sector?}
\author{Jernej F. Kamenik$^{1,2}$ and Christopher Smith$^{3}$\vspace{0.5cm}\\{\small $^{1}$ \textsl{J. Stefan Institute, Jamova 39, P. O. Box 3000, 1001 Ljubljana, Slovenia} }\\{\small $^{2}$ \textsl{Department of Physics, University of Ljubljana, Jadranska 19, 1000 Ljubljana, Slovenia\smallskip} }\\{\small $^{3}$ \textsl{Universit\'{e} Lyon 1 \& CNRS/IN2P3, UMR5822 IPNL, 4 rue Fermi, 69622 Villeurbanne Cedex, France}}}
\date{\today }
\maketitle

\begin{abstract}
The impact a new neutral light particle of spin 0, 1/2, 1, or 3/2 could have on the tiny width of a light Higgs boson is systematically analyzed. To this end, we include all the relevant effective interactions, whether renormalizable or not, and review the possible signatures in the Higgs decay modes with missing energy. This includes the fully invisible Higgs boson decay, as well as modes with SM gauge boson or fermions in the final state. In many cases, simply preventing these modes from being dominant suffices to set tight model-independent constraints on the masses and couplings of the new light states.

\end{abstract}

\section{Introduction}

Recently, some hints of a relatively light Higgs boson were announced by the CMS and Atlas collaborations~\cite{Hints}, with a mass somewhere between 120 and 130 GeV. What may be most spectacular for masses in this range is the tiny width of the Higgs boson in the Standard Model (SM)~\cite{ReviewH}:%
\begin{equation}
\frac{\Gamma_{h}^{SM}}{M_{h}}\approx3\times10^{-5}\;,\label{Eq1}%
\end{equation}
to be compared to $\Gamma_{Z}/M_{Z}\approx\Gamma_{W}/M_{W}\approx2.5\%$ for the SM gauge bosons. So, the Higgs boson would actually be as narrow as the $c\bar{c}$ and $b\bar{b}$ resonances, $\Gamma_{J/\psi}/M_{J/\psi}\approx3\times10^{-5}$ and $\Gamma_{\Upsilon(1S)}/M_{\Upsilon(1S)}\approx0.6\times10^{-5}$~\cite{PDG}.%

\begin{table}[t] \centering
$
\begin{tabular}
[c]{|ccc|ccc|}\hline
\multicolumn{3}{|c|}{Fermions} & \multicolumn{3}{|c|}{Bosons}\\\hline
$M_{H}$ & $120$ & $130$ & $M_{H}$ & $120$ & $130$\\\hline
$b\bar{b}$ & $0.68$ & $0.53$ & $WW$ & $0.13$ & $0.28$\\
$\tau\tau$ & $0.068$ & $0.054$ & $gg$ & $0.068$ & $0.063$\\
$c\bar{c}$ & $0.032$ & $0.025$ & $ZZ$ & $0.015$ & $0.038$\\
$s\bar{s}$ & $0.0006$ & $0.0005$ & $\gamma\gamma$ & $0.0022$ & $0.0022$\\
$\mu\mu$ & $0.0002$ & $0.0002$ & $Z\gamma$ & $0.0011$ & $0.0019$\\\hline
\end{tabular}
\ \ \ $
\caption{Dominant branching ratios of the Higgs boson in the SM (from Ref. \cite{ReviewH}), for a mass of 120 and 130 GeV. The corresponding total widths are $\Gamma_{h}=3.7$ MeV and  $5.0$ MeV. In the $WW$ and $ZZ$ decay channels, one gauge boson is off-shell.}
\label{Hwidth}
\end{table}

With its dramatic suppression in the SM, the Higgs boson width offers an
interesting window for New Physics (NP). Indeed, many models predict the
existence of new light particles (see e.g. Ref.~\cite{ExtraLight} for a
review), for example as the pseudo-Goldstone bosons of some spontaneous
symmetry breaking, or as messengers towards a dark sector to which our world
is only very weakly connected. Provided these particles are neutral,
colorless, and sufficiently weakly interacting, they could have escaped
detection up to now (and will thus be referred to as dark particles). This
also ensures they do not affect the SM Higgs boson production and decay rates
into SM particles, but they could nevertheless open new decay channels.

The dark particles being very weakly interacting, they would show up as missing energy (denoted as $\slashed E$). In particular, they would enhance the invisible width of the Higgs boson, $\Gamma(h^{0}\rightarrow\slashed E)$. But, even if the total width gets enhanced by a factor of ten say, given the current and foreseeable experimental resolutions, it would still be beyond reach (see e.g. Ref.~\cite{Invisible} for a discussion of this issue). Instead, it has to be inferred from the assumed Higgs boson production rate, combined with a measured decay rate in a given SM channel. As said above, though both of these are model-dependent in general, the presence of a dark particle should affect neither of them. Thus, a significant invisible decay rate would systematically suppress the branching ratios for the SM modes (see Table~\ref{Hwidth}), and in particular, no $\gamma\gamma$ signal should have been glimpsed. 

In this way, the tiny Higgs boson width can be measured, and used to constrain the couplings to dark states. Numerically, we take $M_{h}=125$ GeV, and $\Gamma_{h}^{SM}=4$ MeV for definiteness, and will conservatively require the non-standard decay rates to be less than $20\%$ of $\Gamma_{h}^{SM}$~\cite{AtlasInv}. But thanks to Eq.~(\ref{Eq1}), a naive dimensional analysis shows that this should suffice to probe relatively high NP scales:%
\begin{equation}
\frac{1}{5}\times\frac{\Gamma_{h}^{SM}}{M_{h}}\gtrsim\frac{\Gamma_{h}^{dark}%
}{M_{h}}\sim\frac{1}{8\pi}\left(  \frac{M_{h}^{2}}{\Lambda_{d}^{2}}\right)
^{d-4}\;\;\Rightarrow\Lambda_{5}\gtrsim10\;\text{TeV\ ,\ }\Lambda_{6}%
\gtrsim1.1\;\text{TeV\ ,\ }\Lambda_{7}\gtrsim0.5\;\text{TeV\ ,}\label{Reach}%
\end{equation}
where $\Gamma_{h}^{dark}$ is the width of a two-body decay to dark particles induced by an effective operator of dimension $d$. One should also keep in mind the possible improvement at the ILC, where production rates can be much cleaner than $gg\rightarrow h^{0}$, and $\mathcal{B}(h^{0}\rightarrow \slashed E)$ could be measured to a few percent precision~\cite{ILC}.

The possibility to indirectly detect new light particles by measuring the invisible Higgs boson width has already been studied quite extensively. However, most previous works require this new state to be the dark matter candidate. In that case, tight bounds arise from dark matter detection experiments, or WMAP data. In addition, the main focus is in general on the renormalizable couplings to the Higgs boson, the so-called portals~\cite{portals}. For example, the Higgs portal was recently studied in Ref.~\cite{Hportal} and the vector portal in Ref.~\cite{Lebedev11}, while Ref.~\cite{Djouadi11} also included the non-renormalizable couplings to dark fermions.

In the present work, our goal is to study as generically as possible the impact a new light particle could have on the Higgs boson width. So, we will extend previous studies in three directions:

\begin{enumerate}
\item We will not impose any dark matter-based constraints. Indeed, such a new
light state need not be the dark matter candidate, but only has to live long
enough to escape as missing energy at colliders. For instance, it could
originate from some hidden sector, and depending on the dynamics going on
there, may or may not ultimately decay into stable dark matter particles. Our
approach can thus be understood as a model-independent first step towards
unravelling the dark sector dynamics.

\item We will include all the relevant effective operators coupling a dark
particle of spin~0, 1/2, 1, or 3/2 (denoted as $X=\phi,\psi,V,\Psi$) to the
SM, whether renormalizable or not. Since these dark particles have evaded
detection up to now, they have very weak couplings with SM particles. So,
effectively, the dark states can be considered neutral under the full SM gauge
group, and their interactions parametrized by gauge invariant effective
operators. These have recently been constructed in Ref.~\cite{KamenikS11}, and
we will rely on that list quite extensively.

\item We will not only include $h^{0}\rightarrow \slashed E$, but also all the other modes in which the dark states could play a role. Indeed, future bounds on the partially visible branching ratios, with SM particles and missing energy in the final states, may be far better than $\mathcal{B}(h^{0}\rightarrow \slashed E)<20\%$, and would thus, according to Eq.~(\ref{Reach}), probe much higher NP scales. To systematically investigate these signatures, we will retain separately the leading effective interactions involving the Higgs boson alone, together with the SM gauge boson, and together with the SM fermions. In this latter case, we will also distinguish between baryon and lepton number conserving and violating interactions.
\end{enumerate}

Our analysis is organized according to the Higgs boson decay modes. In Section
2, we consider purely invisible decays, $h^{0}\rightarrow\slashed E$ with $\slashed E$ carried away by a pair of dark particles. In Section 3, we turn to modes involving a neutral gauge boson together with one or two dark particles, i.e.
$h^{0}\rightarrow\gamma+\slashed E$ and $h^{0}\rightarrow Z+\slashed E$. These decay channels are further analyzed in Section 4, where the impact of a dark gauge symmetry, and of its breakdown, is studied. In Section 5, we consider decay modes with SM fermions and dark particles in the final state. This includes $h^{0}\rightarrow f\bar{f}+\slashed E$ modes, as well as the invisible and radiative decay processes $h^{0}\rightarrow(\gamma)\psi\nu$ and $h^{0}\rightarrow(\gamma)\Psi\nu$. Finally, our conclusions are presented in Section 6, where we also discuss the interpretation of our results in case the recent hint of a light Higgs boson is not confirmed.

\section{Invisible channels}

Because of the $SU(3)_{C}\otimes SU(2)_{L}\otimes U(1)_{Y}$ invariance, the
simplest way to couple the Higgs boson to a dark SM neutral state is through
the combination%
\begin{equation}
H^{\dagger}H\rightarrow\frac{1}{2}(v^{2}+2vh^{0}+h^{0}h^{0})\;,\label{SSBHH}%
\end{equation}
where the Higgs boson undergoes the shift
\begin{equation}
H\rightarrow\frac{1}{\sqrt{2}}\left(
\begin{array}
[c]{c}%
0\\
v+h^{0}%
\end{array}
\right)  \;,\;\;v\approx246\;\text{GeV\ .}%
\end{equation}
Also, and throughout this work, we assume that the dark scale $\Lambda$ is
larger than the electroweak scale, $\Lambda>v$, so that $H^{\dagger}%
H/\Lambda^{2}\rightarrow v^{2}/(2\Lambda^{2})+...\ $is small and a
perturbative expansion in $1/\Lambda$ is valid.

With this, low-dimensional operators can be constructed for all types of
invisible states~\cite{KamenikS11,Djouadi11}:
\begin{subequations}
\label{HH}%
\begin{align}
\mathcal{H}_{eff}^{0}  &  =\mu^{\prime}H^{\dagger}H\times\phi+\lambda^{\prime
}H^{\dagger}H\times\phi^{\dagger}\phi\;,\label{HHss}\\
\mathcal{H}_{eff}^{1/2}  &  =\frac{c_{LR}}{\Lambda}H^{\dagger}H\times\bar
{\psi}_{L}\psi_{R}+\frac{c_{RL}}{\Lambda}H^{\dagger}H\times\bar{\psi}_{R}%
\psi_{L}\;,\label{HHpp}\\
\mathcal{H}_{eff}^{1}  &  =\varepsilon_{H}H^{\dagger}H\times V_{\mu}V^{\mu
}\;,\label{HHVV}\\
\mathcal{H}_{eff}^{3/2}  &  =\frac{c_{S}}{\Lambda}H^{\dagger}H\times
\overline{\Psi}\hspace{0in}^{\mu}\Psi_{\mu}+\frac{ic_{P}}{\Lambda}H^{\dagger
}H\times\overline{\Psi}\hspace{0in}^{\mu}\gamma_{5}\Psi_{\mu}\;, \label{HHPP}%
\end{align}
\end{subequations}
where all the couplings are real except for $c_{LR}=c_{RL}^{\ast}$, and one should substitute $\psi_{R}\rightarrow\psi_{L}^{C}$ for Majorana fermions. The super-renormalizable $\mu^{\prime}$ and renormalizable $\lambda^{\prime}$ couplings in $\mathcal{H}_{eff}^{0}$ embody the so-called Higgs portal~\cite{Hportal}, while $\varepsilon_{H}$ is part of the vector portal~\cite{Lebedev11}. They are not suppressed by the NP scale $\Lambda$. In the present work, the scalar field is assumed to be charged under some dark quantum number, so that only $\lambda^{\prime}$ occurs.

All these operators induce simultaneously a correction to the invisible particle mass and an invisible $h^{0}\rightarrow \slashed E$ decay mode. The interpretation of this correlation is different for spin~0, $1/2$ and for spin~1, $3/2$ states because for the latter, a mass term explicitly breaks a dark gauge invariance, so we analyze these two cases separately.

\subsection{Spin 0 and 1/2}

The mass corrections and decay rates are simple to get from the effective
interactions:
\begin{subequations}
\label{sspp}%
\begin{align}
\delta m_{\phi}^{2}  &  =\lambda^{\prime}\frac{v^{2}}{2}\;,\;\;\Gamma(h^{0}\rightarrow\phi\phi)=\frac{v^{2}\beta_{\phi}}{8\pi M_{h}}\frac{\lambda^{\prime2}}{2}\;,\\
\delta m_{\psi}  &  =-c_{S}\frac{v^{2}}{2\Lambda}\;,\;\;\Gamma(h^{0}\rightarrow\psi\psi)=\frac{v^{2}\beta_{\psi}}{8\pi M_{h}}(c_{S}^{2}\beta_{\psi}^{2}+c_{P}^{2})\frac{M_{h}^{2}}{\Lambda^{2}}\;,
\end{align}
\end{subequations}
with $\beta_{i}^{2}=1-4r_{i}^{2}$, $r_{i}=m_{i}/M_{h}$, and $c_{S,P}%
=(c_{LR}\pm c_{RL})/2$. The correlation between the mass corrections and the
invisible widths can be interpreted in two ways.

\begin{figure}[t]
\centering         \includegraphics[width=14cm]{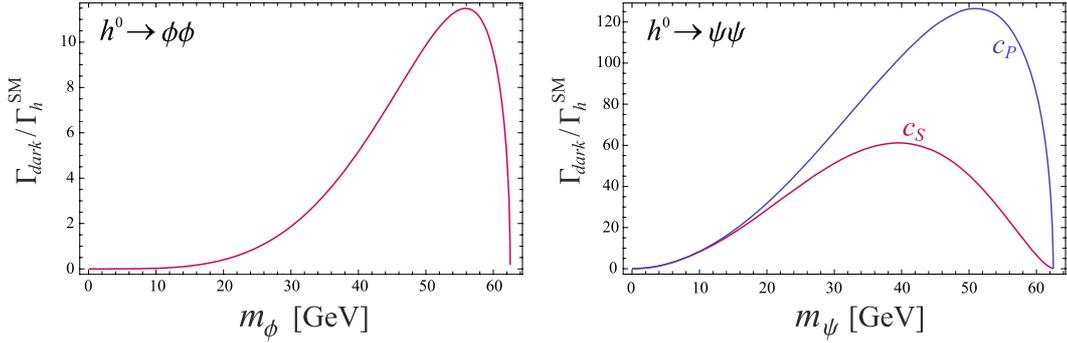}  \caption{Behavior
of $\Gamma(h^{0}\rightarrow\phi\phi)$ [left] and $\Gamma(h^{0}\rightarrow
\psi\psi)$ [right], normalized to the total SM Higgs boson width (set at
$\Gamma_{h}^{SM}=4$ MeV), as a function of $m_{\phi}=\delta m_{\phi}$ and
$m_{\psi}=\delta m_{\psi}$, see Eq.~(\ref{sspp}). For the fermionic case, we
plot separately the scalar ($\sim c_{S}^{2}$) and pseudoscalar ($\sim
c_{P}^{2}$) contributions, even though $c_{LR}$ or $c_{RL}$ is set to one
separately (so $\delta m_{\psi}$ does not vanish).}%
\label{FigHHspsi}%
\end{figure}

\subsubsection*{Upper bound on the dark particle mass}

\begin{figure}[t]
\centering         \includegraphics[width=14cm]{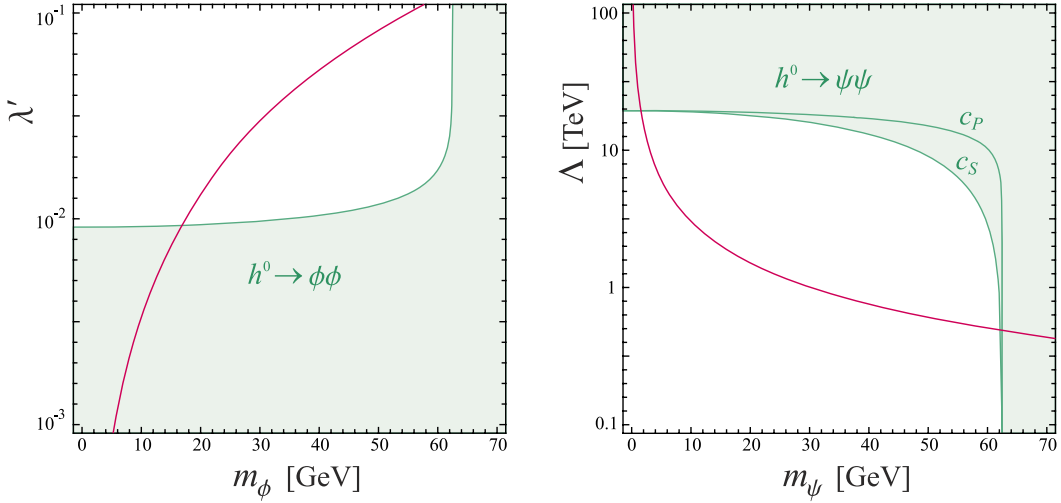}  \caption{Region in the $\lambda^{\prime}-m_{\phi}$ and $\Lambda-m_{\psi}$ plane (setting either $c_{S}$ or $c_{P}$ to $-1$ in the rate, and assuming $\lambda^{\prime}>0$) allowed by the constraint $\Gamma(h^{0}\rightarrow \phi\phi/\psi\psi)<\Gamma_{h}^{SM}/5$ . The red lines correspond to $\bar{m}_{\phi,\psi}=0$, in which case the physical mass is purely electroweak, $m_{\phi,\psi}=\delta m_{\phi,\psi}$, see Eq.~(\ref{sspp}). The upper bounds in Eq.~(\ref{uprBND}) corresponds to the maximal $m_{\phi,\psi}=\delta m_{\phi,\psi}<M_{h}/2$ values such that these lines lie within the allowed regions. The lower bounds in Eq.~(\ref{lwrBND}) are the points at which these curves intersect the boundaries of the allowed regions, since the minimal $m^2_{\phi}= \bar{m}^2_{\phi}+\delta m^2_{\phi}$ and $m_{\psi}= \bar{m}_{\psi}+\delta m_{\psi}$ values such that $\Gamma(h^{0}\rightarrow \phi\phi/\psi\psi)\approx\Gamma_{h}^{SM}/5$ are those for which $\bar{m}_{\phi,\psi}\approx0$ if no fine-tuning between the dark and electroweak mass terms is allowed. Because of the shapes of these curves, the bounds in Eq.~(\ref{uprBND}) and Eq.~(\ref{lwrBND}) coincide.}%
\label{FigHHsspp2}%
\end{figure}

First, if we assume that the invisible states are initially massless or very light, then
\begin{align}
m_{\phi,\psi}\approx\delta m_{\phi,\psi}\;,
\end{align}
with thus $\lambda^{\prime}>0$ and $c_{S}<0$. So, requiring that the invisible $h^{0}$ width does not exceed 20\% of its
predicted SM width translates into upper bounds on the masses of these states,%
\begin{equation}
\left\{\begin{array}[c]{l}%
\Gamma(h^{0}\rightarrow\phi\phi)<\Gamma_{h}^{SM}/5\;\;\Rightarrow m_{\phi}<17\;\text{GeV}\;,\\
\Gamma(h^{0}\rightarrow\psi\psi)<\Gamma_{h}^{SM}/5\;\;\Rightarrow m_{\psi}<1.6\;\text{GeV}\;,
\end{array}\right. \label{uprBND}
\end{equation}
but for a small range close to the kinematical threshold $m_{\phi,\psi}\approx M_{h}/2$, see Fig.~\ref{FigHHspsi}. In the fermionic case, $c_{LR}$ or $c_{RL}$ are set to one separately, leading to similar bounds. Clearly, the constraint is much tighter for dark fermions, because of the extra $M_{h}^{2}$ power occurring in $\Gamma(h^{0}\rightarrow\psi\psi)$. Such a low $m_{\psi}$ makes it accessible to rare $B$ decays. In this respect, the bound $\Gamma(h^{0}\rightarrow\phi\phi/\psi\psi)<\Gamma_{h}^{SM}/5$ translates as (see Fig.~\ref{FigHHsspp2}) 
\begin{equation}
\left\{\begin{array}[c]{l}%
\Gamma(h^{0}\rightarrow\phi\phi)<\Gamma_{h}^{SM}/5\;\;\Rightarrow\lambda^{\prime}<0.01\;,\\
\Gamma(h^{0}\rightarrow\psi\psi)<\Gamma_{h}^{SM}/5\;\;\Rightarrow\Lambda\gtrsim20\;\text{TeV}\;,
\end{array}\right.
\end{equation}
for $m_{\phi,\psi}\lesssim30$ GeV (above which the phase-space suppression kicks in). These values are too small to lead to sizeable flavor-changing Higgs penguins, and thus an impact on rare $B$ decays would require the presence of direct couplings to quark fields. Note, by the way, that the scale reached for the dark fermions is similar to the scales accessible using rare $B$ and $K$ decays~\cite{KamenikS11}.

\subsubsection*{Lower bound on the dark particle mass}

A second way to view the correlation in Eq.~(\ref{sspp}) starts by imagining that a 20\% enhancement of the $h^{0}$ width over the SM value is observed. Then, if it is due to the coupling to the invisible states, $\Gamma(h^{0}\rightarrow\phi\phi/\psi
\psi)\approx\Gamma_{h}^{SM}/5$, one gets a lower bound on the physical mass $m_{\phi,\psi}$ of the invisible states (see Fig.~\ref{FigHHsspp2})
\begin{equation}
\left\{\begin{array}[c]{l}%
\Gamma(h^{0}\rightarrow\phi\phi)\approx\Gamma_{h}^{SM}/5\;\;\Rightarrow m_{\phi}\gtrsim17\;\text{GeV}\;,\\
\Gamma(h^{0}\rightarrow\psi\psi)\approx\Gamma_{h}^{SM}/5\;\;\Rightarrow m_{\psi}\gtrsim1.6\;\text{GeV}\;.
\end{array} \right. \label{lwrBND}
\end{equation}
Indeed, it is highly unlikely that the Lagrangian mass $\bar{m}_{\phi,\psi}$ of these states, originating from the dark sector dynamics, is in any way related to the electroweak symmetry breaking. Hence, with $m_{\phi}^{2}=\bar{m}_{\phi}^{2}+\delta m_{\phi}^{2}$, $m_{\psi}=\bar{m}_{\psi}+\delta m_{\psi}$, and forbidding a strong cancellation between $\bar{m}_{\phi,\psi}$ and $\delta m_{\phi,\psi}$, the correction $\delta m_{\phi,\psi}$ acts as a lower bound for $m_{\phi,\psi}$.

\subsection{Spin 1 and 3/2}

The Higgs boson couplings to the spin~1 and~$3/2$ fields in Eq.~(\ref{HH}) explicitly break the gauge
invariance of their respective free massless Lagrangian. Hence, when computing the $h^{0}\rightarrow VV$ ($h^{0}\rightarrow\Psi\Psi$) rate, the $1/m_{V}$ ($1/m_{\Psi}$) term of the polarization (spin) sum is not projected out:
\begin{subequations}
\label{VVPP}%
\begin{align}
\delta m_{V}^{2} &  =\varepsilon_{H}v^{2}\;,\;\;\Gamma(h^{0}\rightarrow
VV)=\frac{1}{2}\frac{v^{2}\beta_{V}}{8\pi M_{h}}\frac{\varepsilon_{H}^{2}}%
{2}\frac{3-2\beta_{V}^{2}+3\beta_{V}^{4}}{4r_{V}^{4}}\;,\\
\delta m_{\Psi} &  =c_{S}\frac{v^{2}}{2\Lambda}\;,\;\;\Gamma(h^{0}%
\rightarrow\Psi\Psi)=\frac{v^{2}\beta_{\Psi}}{8\pi M_{h}}\frac{c_{S}^{2}%
\beta_{\Psi}^{2}\beta_{\Psi}^{\prime}+c_{P}^{2}\beta_{\Psi}^{\prime\prime}%
}{9r_{\Psi}^{4}}\frac{M_{H}^{2}}{\Lambda^{2}}\;,
\end{align}
with $\beta_{\Psi}^{\prime}=(5-6\beta_{\Psi}^{2}+9\beta_{\Psi}^{4})/8$ and
$\beta_{\Psi}^{\prime\prime}=(9-6\beta_{\Psi}^{2}+5\beta_{\Psi}^{4})/8$,
including the factor~$1/2$ for the identical dark vectors. The physical masses
are $m_{V}^{2}=\bar{m}_{V}^{2}+\delta m_{V}^{2}$ and $m_{\Psi}=\bar{m}_{\Psi
}+\delta m_{\Psi}$, with $\bar{m}_{V,\Psi}$ the Lagrangian mass parameters
(for a complex spin~3/2 field).

If the dark gauge invariance is dominantly broken by the $H^{\dagger}H$
operators, the masses of $V$ and $\Psi$ are set by the electroweak spontaneous
symmetry breaking at $m_{V,\Psi}=\delta m_{V,\Psi}$ (which requires
$\varepsilon_{H}>0$ and $c_{S}>0$). They are thus directly related to the
corresponding invisible Higgs boson decay rates. But, because of the
enhancement due to the $1/m_{V}$ or $1/m_{\Psi}$ terms mentioned above, simply
requiring the invisible width not to exceed $\Gamma_{h}^{SM}$ pushes these
masses at the kinematical boundary%
\end{subequations}
\begin{equation}
\left\{ \begin{array}[c]{l}%
\Gamma(h^{0}\rightarrow VV)<\Gamma_{h}^{SM}\;\;\overset{\bar{m}_{V}=0}{\Rightarrow}m_{V}\gtrsim M_{h}/2\;,\\
\Gamma(h^{0}\rightarrow\Psi\Psi)<\Gamma_{h}^{SM}\;\;\overset{\bar{m}_{\Psi}=0}{\Rightarrow}m_{\Psi}\gtrsim M_{h}/2\;.
\end{array}\right.  \label{HHVPbnd}%
\end{equation}
This result crucially depends on the smallness of $\Gamma_{h}^{SM}$. For example, the mass of the dark vector would be left unconstrained if $\Gamma_{h}\gtrsim320$ MeV for $M_{h}=125$ GeV. This can be understood by setting $m_{V}^{2}=\varepsilon_{H}v^{2}$ in $\Gamma(h^{0}\rightarrow VV)$, and then taking the limit $\varepsilon_{H}\rightarrow0$:%
\begin{equation}
\Gamma(h^{0}\rightarrow VV)\overset{\varepsilon_{H}\rightarrow0}{=}\frac{M_{h}^{3}}{32\pi v^{2}}\approx320\;\text{MeV}\;.\label{Rate0a}%
\end{equation}
Since this is about 80 times larger than the total SM width, the presence of this invisible decay mode reduces all the SM branching ratios by about 80. Obviously, the recent hint of a Higgs boson in the $\gamma\gamma$ channel, if confirmed, would rule out such a systematic suppression. For spin~3/2 final states, the rate diverges if either $\Lambda\rightarrow\infty$ or $c_{S}\rightarrow0$ because the mass is only linearly dependent on $c_{S}/\Lambda$.

\begin{figure}[t]
\centering         \includegraphics[width=14cm]{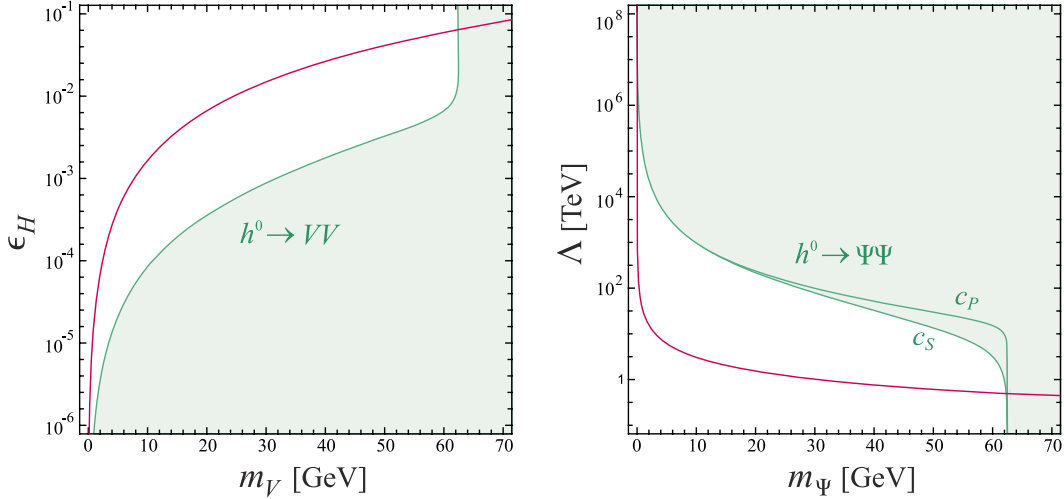}  \caption{Region in the $\varepsilon_{H}-m_{V}$ and $\Lambda-m_{\Psi}$ plane (setting either $c_{S}$ or $c_{P}$ to one in the rate, and assuming $\varepsilon_{H}>0$) allowed by the constraint $\Gamma(h^{0}\rightarrow VV/\Psi\Psi)<\Gamma_{h}^{SM}/5$ . The red lines correspond to $\bar{m}_{V,\Psi}=0$, in which case the physical mass is purely electroweak, $m_{V,\Psi}=\delta m_{V,\Psi}$, see Eq.~(\ref{VVPP}). Except above about $M_{h}/2$, these lines lie entirely outside of the allowed region (compare with Fig.~\ref{FigHHsspp2}), corresponding to the bounds in Eq.~(\ref{HHVPbnd}). Note that the mass values to the left of these curves require a fine-tuning between the dark and electroweak contributions to the physical masses.}%
\label{FigHHVPsi}%
\end{figure}

Of course, once we allow for the operators in Eq.~(\ref{HHVV}, \ref{HHPP}), there is a priori no reason for $\bar{m}_{V,\Psi}$ to vanish since the dark gauge symmetry is explicitly broken. But, with a sufficiently large$\ \bar{m}_{V,\Psi}$, the limits~(\ref{HHVPbnd}) can be evaded thanks to the softening of the $1/m_{V,\Psi}^{4}$ singularity. What we can nevertheless say is that the bounds from $\Gamma_{h}^{SM}$ ask for $|\delta m_{V}|<m_{V}$ and $|\delta m_{\Psi}|<m_{\Psi}$, thereby excluding large portions of the $\varepsilon_{H}-m_{V}$ and $\Lambda-m_{\Psi}$ planes, see Fig.~\ref{FigHHVPsi}. It should be noted that compared to the kinetic mixing~\cite{Kinetic}, $\chi B_{\mu\nu}V^{\mu\nu}$, the exclusion region in the $\varepsilon_{H}-m_{V}$ plane is larger than in the corresponding $\chi-m_{V}$ plane, see Refs.~\cite{Williams,DarkVector,KamenikS11}.

In general, some dynamical information on the dark sector is needed to draw a definite conclusion about the physical mass of the dark state. In addition, compared to the spin~0 and 1/2 cases, a direct link between $\delta m_{V,\Psi}$ and $\bar{m}_{V,\Psi}$ is quite possible, since both arise from a breaking of the dark gauge symmetry (this will be further studied in Section~4). For example, if we require $\bar{m}_{V}^{2}=\varepsilon_{H}v_{dark}^{2}$, so that $m_{V}^{2}=\varepsilon_{H}(v^{2}+v_{dark}^{2})$, then
\begin{equation}
\Gamma(h^{0}\rightarrow VV)<\Gamma_{h}^{SM}/5\;\;\Rightarrow v_{dark}%
>1.1\;\text{TeV },
\end{equation}
for $m_{V}\ll M_{h}$. Similarly, for the spin~3/2 case, let us recall the supergravity~\cite{Nilles} relation $M_{Planck}=\Lambda_{SUSY}^{2}/(\sqrt{3}\bar{m}_{\Psi})=(8\pi G_{N})^{-1/2}$. Upon identifying $\Lambda_{SUSY}$ with the dark symmetry breaking scale $v_{dark}$, and $M_{Planck}$ with $\Lambda/c_{S}$, this suggests setting $\Lambda=c_{S}(v^{2}+v_{dark}^{2})/(2m_{\Psi})$. The divergence as $m_{\Psi}\rightarrow0$ is not cured by this replacement, so the bound on $v_{dark}$ strongly depends on $m_{\Psi}$,%
\begin{equation}
\Gamma(h^{0}\rightarrow\Psi\Psi)<\Gamma_{h}^{SM}/5\;\;\Rightarrow v_{dark}\gtrsim(2,\;4,\;14,\;45)\;\text{TeV\ for }m_{\Psi}\approx (30,\;10,\;1,\;0.1)\;\text{GeV\ },
\end{equation}
when $c_{P,S}$ are of $\mathcal{O}(1)$.

\section{Gauge channels}

\begin{figure}[t]
\centering   \includegraphics[width=12cm]{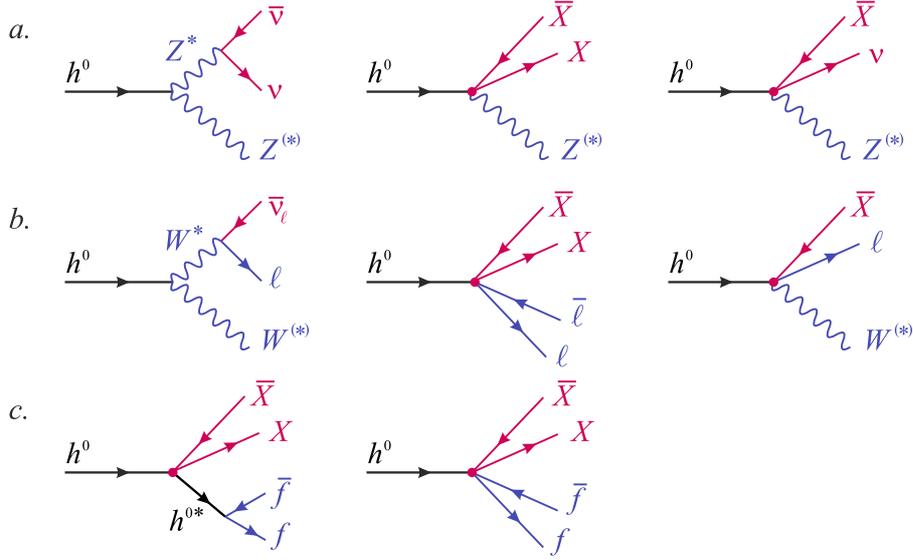}  \caption{($a$) Comparison between the SM processes induced by $h^{0}\rightarrow ZZ^{\ast}[\rightarrow\nu\bar{\nu}]$ and the contact interactions producing $h^{0}\rightarrow ZXX$, $X=\phi,\psi,V,\Psi$ or $h^{0}\rightarrow Z\nu X$, $X=\psi,\Psi$. Generically, the NP processes are suppressed by $O(v^{2}%
/\Lambda^{2})$ since the $Z$ propagator of the SM process gets substituted by a $1/\Lambda^{2}$ factor for the effective operators. ($b$) Same for the SM $h^{0}\rightarrow W^{(\ast)}W^{\ast}$ processes. ($c$) The fermionic channels induced by the operator of Eq.~(\ref{HH}), and enhanced by $\mathcal{O}(M_{h}^{2}/\Lambda^{2})$ compared to the dark effective interactions of Eq.~(\ref{HFermion}) when MFV holds.}%
\label{SMvsNP}%
\end{figure}

The invisible states can also be produced in $h^{0}$ decays in conjunction with SM particles. The simplest effective interactions generating such final states are built from the Higgs vector current ($c_{W}\equiv\cos\theta_{W}$)%
\begin{equation}
H^{\dagger}\overleftrightarrow{\mathcal{D}}\hspace{0in}^{\mu}H\equiv H^{\dagger}\overleftarrow{\mathcal{D}}\hspace{0in}^{\mu}H-H^{\dagger}\overrightarrow{\mathcal{D}}\hspace{0in}^{\mu}H\rightarrow\frac{ig}{2c_{W}}(v^{2}+2vh^{0}+h^{0}h^{0})Z^{\mu}\;,\label{HDH}%
\end{equation}
and thus naturally produce decay channels with an external $Z$ boson. For each type of dark particle, the leading operators are
\begin{subequations}
\label{GaugeOp}%
\begin{align}
\mathcal{H}_{eff}^{0} &  =\frac{c_{\phi}^{H}}{\Lambda^{2}}H^{\dagger}\overleftrightarrow{\mathcal{D}}\hspace{0in}_{\mu}H\times\phi^{\dagger}\overleftrightarrow{\partial}\hspace{0in}^{\mu}\phi\;,\label{GaugeOp1}\\
\mathcal{H}_{eff}^{1/2} &  =\frac{c_{L}^{H}}{\Lambda^{2}}iH^{\dagger}\overleftrightarrow{\mathcal{D}}\hspace{0in}_{\mu}H\times\bar{\psi}_{L}\gamma^{\mu}\psi_{L}+\frac{c_{R}^{H}}{\Lambda^{2}}iH^{\dagger}\overleftrightarrow{\mathcal{D}}\hspace{0in}_{\mu}H\times\bar{\psi}_{R}\gamma^{\mu}\psi_{R}\;,\label{GaugeOp2}\\
\mathcal{H}_{eff}^{1} &  =\varepsilon_{2}^{H}iH^{\dagger}\overleftrightarrow{\mathcal{D}}\hspace{0in}_{\mu}H\times V^{\mu}\;,\label{GaugeOp3}\\
\mathcal{H}_{eff}^{3/2} &  =\frac{c_{V}^{H}}{\Lambda^{2}}iH^{\dagger}\overleftrightarrow{\mathcal{D}}\hspace{0in}_{\mu}H\times\overline{\Psi}\hspace{0in}^{\rho}\gamma^{\mu}\Psi_{\rho}+\frac{c_{A}^{H}}{\Lambda^{2}}iH^{\dagger}\overleftrightarrow{\mathcal{D}}\hspace{0in}_{\mu}H\times\overline{\Psi}\hspace{0in}^{\rho}\gamma^{\mu}\gamma_{5}\Psi_{\rho}\;.\label{GaugeOp4}%
\end{align}

Actually, none of the dimension six operators could have a visible impact on $\Gamma_{h}$. First, note that these operators induce $h^{0}\rightarrow Z+\slashed E$ decay modes, and are thus experimentally entangled with $h^{0}\rightarrow ZZ^{\ast}[\rightarrow\nu\bar{\nu}]$. In the SM, the amplitude for this process is (see Fig.~\ref{SMvsNP}$a$)%
\end{subequations}
\begin{equation}
\mathcal{M}(h^{0}\rightarrow Z\nu\bar{\nu})=i\frac{2M_{Z}^{2}}{v}g^{\mu\alpha
}\frac{g_{\alpha\nu}}{T^{2}-M_{Z}^{2}}\frac{g}{2c_{W}}\{\bar{\nu}_{L}%
\gamma^{\nu}\nu_{L}\}\varepsilon_{\mu}^{\ast}\approx i\frac{g}{vc_{W}}%
\{\bar{\nu}_{L}\gamma^{\mu}\nu_{L}\}\varepsilon_{\mu}^{\ast}\;,\label{SMampli}%
\end{equation}
where $T^{2}=(p_{\nu}+p_{\bar{\nu}})^{2}$ is the virtual $Z$ boson momentum, and $\varepsilon_{\mu}^{\ast}$ the on-shell $Z$ boson polarization vector. Because both $Z$ bosons cannot be on-shell simultaneously for $M_{h}\approx125$ GeV, this amplitude can be matched onto a dimension-six operator of the same form as in Eq.~(\ref{GaugeOp}), but with an $\mathcal{O}(1)$ Wilson coefficient and a scale $\Lambda_{SM}\approx v$. With $\Lambda>v$ for the dark dimension-six operators, the $h^{0}\rightarrow Z\phi\phi$, $Z\psi\psi$, or $Z\Psi\Psi$ processes could at best slightly enhance the $h^{0}\rightarrow Z+\slashed E$ channel\footnote{For the spin 3/2 operators, the singularity due to the spin sum is cured by setting $\Lambda=v^{2}/(2m_{\Psi})$. Though the $h^{0}\rightarrow Z\Psi\Psi$ rate is then no longer directly suppressed by $\Lambda$, it is suppressed by $(M_{h}/v)^{6}$ as well as by its reduced phase space, and $\mathcal{B}(h^{0}\rightarrow Z\Psi\Psi)\lesssim10^{-6}$. Note also that the $c_{V}^{H}$ operator is absent if $\Psi$ obeys the Majorana condition.}, which is itself a tiny fraction of $\Gamma_{h}^{SM}$ (see Table~\ref{Hwidth}). So, with a precise measurement of $h^{0}\rightarrow Z\nu\bar{\nu}$ beyond our reach experimentally, and the impact on $\Gamma_{h}$ far below the percent level, the presence of an SM gauge boson in the final state does not open interesting windows for scalar or fermionic dark states.

For the production of dark vector states, the simplest operator is renormalizable. Its most interesting feature is to force any possible dark gauge symmetry to be broken along with the electroweak symmetry. As a result, the $Z^{\mu}V_{\mu}$ mixing induces a correction to the $V$ mass and the $1/m_{V}^{2}$ term of the polarization sum is not projected out in the
$h^{0}\rightarrow ZV$ rate:%
\begin{equation}
\delta m_{V}^{2}=-(\varepsilon_{2}^{H})^{2}\frac{g^{2}v^{4}}{4M_{Z}^{2}%
c_{W}^{2}}=-(\varepsilon_{2}^{H})^{2}v^{2}\;,\;\;\Gamma(h^{0}\rightarrow
ZV)=(\varepsilon_{2}^{H})^{2}\frac{g^{2}v^{2}\sqrt{\lambda}}{64\pi M_{h}%
c_{W}^{2}}\frac{12r_{Z}^{2}r_{V}^{2}+\lambda}{r_{Z}^{2}r_{V}^{2}}\;,
\label{HZV}%
\end{equation}
where $\lambda=\lambda(1,r_{Z}^{2},r_{V}^{2})$ and $\lambda(a,b,c)=a^{2}%
+b^{2}+c^{2}-2(ab+ac+bc)$. Because the mass correction is necessarily
negative, we have to set $\bar{m}_{V}>0$. As a first step, let us assume that
$m_{V}^{2}=\bar{m}_{V}^{2}+\delta m_{V}^{2}\approx-\delta m_{V}^{2}$. This is
the smallest $m_{V}$ compatible with the absence of fine-tuning between the
two terms. Then, the unknown $\varepsilon_{2}^{H}$ coupling can be eliminated
in favor of $m_{V}$, thereby ensuring a safe behavior for the rate as the
vector gets light
\begin{equation}
\Gamma(h^{0}\rightarrow ZV)=\frac{M_{h}g^{2}\sqrt{\lambda}}{64\pi c_{W}^{2}%
}\frac{12r_{V}^{2}r_{Z}^{2}+\lambda}{r_{Z}^{2}}\overset{m_{V}\rightarrow0}%
{=}\frac{g^{2}M_{h}}{64\pi c_{W}^{2}}\frac{(1-r_{Z}^{2})^{3}}{r_{Z}^{2}%
}\approx66\;\text{MeV}\;. \label{Rate0b}%
\end{equation}
This is more than 15 times the total SM width, so we can safely set the bound%
\begin{equation}
\Gamma(h^{0}\rightarrow ZV)<\Gamma_{h}^{SM}\;\;\Rightarrow m_{V}\gtrsim
M_{h}-M_{Z}\;. \label{BND1}%
\end{equation}
At the same time, when $\varepsilon_{2}^{H}$ is not tiny, the $ZV$ mixing couples $V$ to the SM $Z$ matter current. In that case, there are already tight constraints from various low-energy observables. For our purpose, it suffices to note that this mixing also shifts the $Z$ mass, and hence the electroweak $\rho$ parameter, as
\begin{equation}
\delta\rho=\frac{\delta m_{V}^{2}}{M_{Z}^{2}}\;\overset{m_{V}^{2}%
\approx-\delta m_{V}^{2}}{\rightarrow}-\frac{m_{V}^{2}}{M_{Z}^{2}%
}\;\;\Rightarrow m_{V}<2.4\;\text{GeV\ ,} \label{BND2}%
\end{equation}
where we use $\rho=1.0004_{-0.0011}^{+0.0029}$ at 2$\sigma$~\cite{PDG}. So, the window is completely closed: the tiny SM width $\Gamma_{h}^{SM}$ combined with the $\rho$ parameter forbid the presence of a light vector state coupled to the SM particles through the $iH^{\dagger}\overleftrightarrow{\mathcal{D}}\hspace{0in}_{\mu}H\times V^{\mu}$ operator when its mass is of the order of $m_{V}^{2}\approx-\delta m_{V}^{2}$.

\begin{figure}[t]
\centering        \includegraphics[width=14cm]{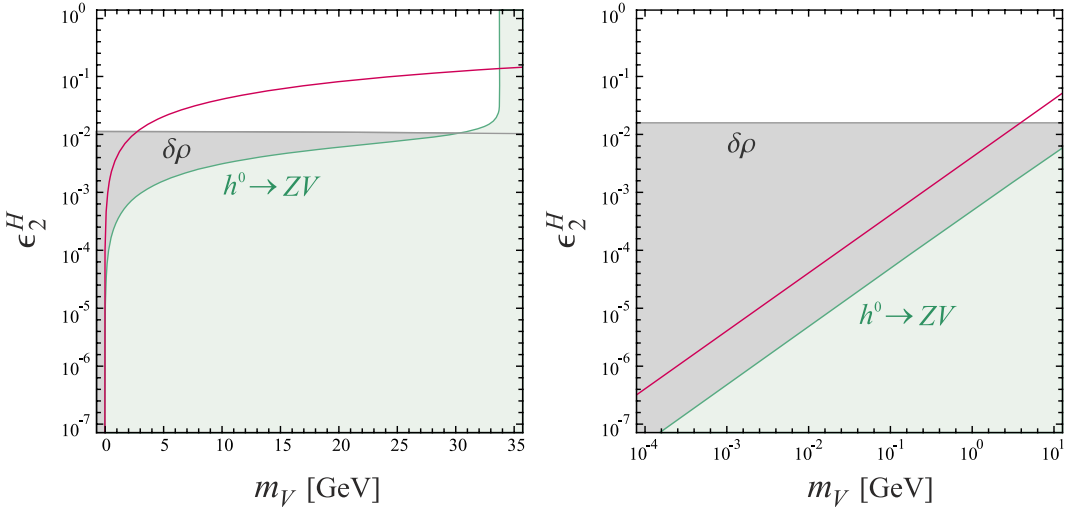}  \caption{Region in the $\varepsilon_{2}^{H}-m_{V}$ plane allowed by the constraint $\Gamma(h^{0}\rightarrow VZ)<\Gamma_{h}^{SM}/5$ [green] and by the electroweak $\rho$ parameter [grey]. The red line corresponds to $m_{V}^{2}=-\delta m_{V}^{2}$, see Eq.~(\ref{HZV}). Except above about $M_{h}-M_{Z}$, it lies entirely outside of the green region, corresponding to the bound~(\ref{BND1}), and enters the grey area at around $m_{V}\approx2.4$ GeV, see Eq.~(\ref{BND2}). The $m_{V}$ values to the left of this curve require a fine-tuning between $\bar{m}_{V}^{2}$ and $\delta m_{V}^{2}$. The right panel shows the same using a logarithmic mass scale, in order to emphasize the strengthening of the bound as $m_{V}\rightarrow0$. Note that the constraint on $\varepsilon_{2}^{H}$ from the $\rho$ parameter depends on $m_{V}$, but this is hidden by the plot scales.}%
\label{FigHHVPsiGauge}%
\end{figure}

If we relax the constraint $m_{V}^{2}\approx-\delta m_{V}^{2}$, the bound from
$\Gamma_{h}^{SM}$ translates as an exclusion region in the $\varepsilon
_{2}^{H}-m_{V}$ plane, see Fig.~\ref{FigHHVPsiGauge}. Given that the
$\Gamma(h^{0}\rightarrow ZV)$ rate in Eq.~(\ref{Rate0b}) is closer to
$\Gamma_{h}^{SM}$ than $\Gamma(h^{0}\rightarrow VV)$, Eq.~(\ref{Rate0a}), this
exclusion region is less extensive than that shown in Fig.~\ref{FigHHVPsi}.
Still, this mode should not be disregarded, because it may be easier to
measure $h^{0}\rightarrow ZV$ than $h^{0}\rightarrow VV$. So in the future,
this channel may provide a powerful window for dark vector states.

\section{Dark gauge connection between the invisible and gauge channels}

At this stage, it is worth discussing in more details the properties of the
$H^{\dagger}H\times V_{\mu}V^{\mu}$ and $iH^{\dagger}\overleftrightarrow
{\mathcal{D}}\hspace{0in}_{\mu}H\times V^{\mu}$ couplings. At first sight, there is a
major difference between the two. The $H^{\dagger}H\times V_{\mu}V^{\mu}$
coupling manifestly breaks the dark gauge symmetry, no matter what happens to
the Higgs field (but for $H=0$). So, there is no reason for the bare mass
$\bar{m}_{V}$ to be absent, or related to the electroweak scale in any way.
The $iH^{\dagger}\overleftrightarrow{\mathcal{D}}\hspace{0in}_{\mu}H\times V^{\mu}$ only
breaks the dark gauge symmetry through the specific dynamics of the Higgs
field in the SM. Indeed, before the electroweak symmetry breaking,
$\mathcal{D}^{2}H^{\dagger}H-H^{\dagger}\mathcal{D}^{2}H=0$ at the classical level for a Higgs
doublet only coupled to the SM gauge interactions~\cite{BuchmullerW86}. So it
looks like the need for $\bar{m}_{V}>0$ arises only after the electroweak
symmetry breaking. The constraint $m_{V}^{2}\approx|\delta m_{V}^{2}|$ would
thus appear natural, but is nevertheless excluded experimentally.

\subsection{Higgs couplings}

To gain some insight into the gauge properties of the $H^{\dagger}H\times
V_{\mu}V^{\mu}$ and $iH^{\dagger}\overleftrightarrow{\mathcal{D}}\hspace{0in}_{\mu
}H\times V^{\mu}$ operators, let us couple the dark vector to the SM by
assigning a dark $U(1)$ charge to the Higgs doublet,
\begin{equation}
\bar{\mathcal{D}}^{\mu}H=\left(  \mathcal{D}^{\mu}-i\frac{\lambda}{2}V^{\mu}\right)  H=\left(
\partial^{\mu}-ig\frac{\tau^{3}}{2}W_{3}^{\mu}-i\frac{g^{\prime}}{2}B^{\mu
}-i\frac{\lambda}{2}V^{\mu}\right)  H\;. \label{HiggsGI}%
\end{equation}
(for simplicity, we do not include the $W^{\pm}$ in the covariant derivative).
Then, the Higgs doublet kinetic term becomes:%
\begin{equation}
\mathcal{L}_{Higgs}^{1}=\bar{\mathcal{D}}_{\mu}H^{\dagger}\bar{\mathcal{D}}^{\mu}H=\mathcal{D}_{\mu
}H^{\dagger}\mathcal{D}^{\mu}H-i\frac{\lambda}{2}H^{\dagger}\overleftrightarrow
{\mathcal{D}}\hspace{0in}_{\mu}H\times V^{\mu}+\frac{\lambda^{2}}{4}H^{\dagger}H\times
V_{\mu}V^{\mu}\;.
\end{equation}
In this case, both the $H^{\dagger}H\times V_{\mu}V^{\mu}$ and $iH^{\dagger
}\overleftrightarrow{\mathcal{D}}\hspace{0in}_{\mu}H\times V^{\mu}$ couplings are needed
to ensure the dark gauge invariance, and their strengths have to be related.
Of course, since SM fermions are not charged under the dark $U(1)$, this
symmetry is explicitly broken by the Yukawa couplings (for the same reason,
$\mathcal{D}^{2}H^{\dagger}H-H^{\dagger}\mathcal{D}^{2}H\neq0$ in the SM~\cite{BuchmullerW86}),
but for now, let us concentrate on the gauge and Higgs sectors.

After the electroweak symmetry breaking, the mass term is not diagonal%
\begin{equation}
\bar{\mathcal{D}}_{\mu}H^{\dagger}\bar{\mathcal{D}}^{\mu}H\rightarrow\frac{1}{2}\partial_{\mu
}h^{0}\partial^{\mu}h^{0}+\frac{M_{Z}^{2}}{2}\left(  1+\frac{h^{0}}{v}\right)
^{2}\left(  \bar{Z}_{\mu}+\bar{\lambda}\bar{V}_{\mu}\right)  \left(  \bar
{Z}^{\mu}+\bar{\lambda}\bar{V}^{\mu}\right)  \;,\;\bar{\lambda}=\frac{\lambda
}{2M_{Z}/v}\;,\label{Zcoupl}%
\end{equation}
so the gauge bosons need to be redefined (from now on, the unrotated
Lagrangian fields are denoted with a bar). The physical $Z$ mass is corrected
as $M_{Z}^{2}(1+\bar{\lambda}^{2})$, with $M_{Z}^{2}c_{W}^{2}=M_{W}^{2}$ and
$M_{W}=gv/2$, but the physical $V$ state remains massless because the
correction coming from the $\bar{Z}_{\mu}\times\bar{V}^{\mu}$ mixing cancels
exactly that from the $H^{\dagger}H\times\bar{V}_{\mu}\bar{V}^{\mu}$ term.
Further, diagonalizing the gauge bosons' mass term diagonalizes also their couplings
to the Higgs boson, so no direct $h^{0}ZV$ or $h^{0}VV$ vertices remain.

On the contrary, if a bare mass $\bar{m}_{V}$ is introduced, the vector boson
masses and couplings to the Higgs boson get misaligned. In the mass eigenstate
basis, the Higgs couplings to $V$ no longer vanish, but are proportional to
its physical mass $m_{V}$:%
\begin{align}
\bar{\mathcal{D}}_{\mu}H^{\dagger}\bar{\mathcal{D}}^{\mu}H+\frac{1}{2}\bar{m}_{V}^{2}\bar{V}_{\mu
}\bar{V}^{\mu} &  \rightarrow\frac{1}{2}\partial_{\mu}h^{0}\partial^{\mu}%
h^{0}+\frac{M_{Z}^{2}}{2}\left(  1+\frac{h^{0}}{v}\right)  ^{2}\left(
1+\bar{\lambda}^{2}\right)  Z_{\mu}Z^{\mu}\nonumber\\
&  \;\;\;\;+\frac{m_{V}^{2}}{2}\left(  \frac{2h^{0}}{v}+\frac{h^{0}h^{0}%
}{v^{2}}\right)  \left(  2\bar{\lambda}Z_{\mu}V^{\mu}+\bar{\lambda}^{2}%
r_{VZ}^{2}V_{\mu}V^{\mu}\right)  +\frac{m_{V}^{2}}{2}V_{\mu}V^{\mu}\;,
\end{align}
where we retain only the leading order in $\bar{\lambda}$ and $r_{VZ}\equiv
m_{V}/M_{Z}$. Because of these mass factors, not only are the $h^{0}%
\rightarrow ZV$ and $h^{0}\rightarrow VV$ rates safe in the $m_{V}%
\rightarrow0$ limit, but they actually vanish:
\begin{subequations}
\label{RateGauge}%
\begin{align}
\Gamma(h^{0} &  \rightarrow VV)=\frac{M_{h}^{3}\beta_{V}}{128\pi v^{2}}%
r_{VZ}^{4}\left(  3-2\beta_{V}^{2}+3\beta_{V}^{4}\right)  \bar{\lambda}%
^{4}\;\;\overset{m_{V}=57\;\text{GeV}}{\approx}(14\;\text{MeV)}\times
\bar{\lambda}^{4}\;,\\
\Gamma(h^{0} &  \rightarrow ZV)=\frac{M_{h}^{3}\sqrt{\lambda}}{16\pi v^{2}%
}r_{VZ}^{2}\left(  12r_{V}^{2}r_{Z}^{2}+\lambda\right)  \bar{\lambda}%
^{2}\;\;\overset{m_{V}=29\;\text{GeV}}{\approx}(6\;\text{MeV)}\times
\bar{\lambda}^{2}\;\;.
\end{align}
Impressively, this occurs even when the Higgs mechanism is not generating the bulk of the mass of the dark vector, since a non-zero $\bar{m}_{V}$ must arise from the dark sector. Phenomenologically, for the $m_{V}$ values which maximize each rate, requiring $\Gamma(h^{0}\rightarrow VV/ZV)<\Gamma_{h}^{SM}/5$ translates as $\bar{\lambda}\lesssim0.5$. For comparison, the $\rho$ parameter is modified as $\delta\rho\approx-\bar{\lambda}^{2}$, which translates~\cite{PDG} as $\bar{\lambda}<0.03$. Other electroweak constraints may even be stronger (see e.g. Ref.~\cite{Williams,DarkVector,Marciano12}), but that from the $\rho$ parameter already suffices to push the branching ratios down to $\mathcal{B}(h^{0}\rightarrow VV)\lesssim3\times10^{-6}$ and $\mathcal{B}(h^{0}\rightarrow ZV)\lesssim1.4\times10^{-3}$. So, even though $\Gamma(h^{0}\rightarrow ZV)$ may not be particularly small (it could be of the same order as $\Gamma(h^{0}\rightarrow\gamma\gamma)$), these processes cannot significantly affect $\Gamma_{h}^{SM}$, no matter the vector mass.

So, whether $\bar{m}_{V}$ vanishes or not, the dark vector may be best searched for through its couplings to SM fermions. As said earlier, given Eq.~(\ref{HiggsGI}), the Yukawa couplings are invariant under the dark $U(1)$ only once fermions are charged. But, assigning the adequate dark charges to the fermions, the $V$ field couples exactly like the $B$ field, so a unitary rotation permits one to completely decouple $V$ from the SM. Of course, this unitary rotation can be performed only when $\bar{m}_{V}=0$. If $\bar{m}_{V}>0$, the mass matrix selects the physical eigenstates, and the $V$ field retains some couplings to the fermions, tuned by $\lambda$. In this respect, note that $\lambda$ should be small for $V$ to show up as missing energy, and not as a fermion pair, at colliders. In addition, there are already very tight constraints on $\bar{\lambda}$ from low-energy experiments (from beam dump, anomalous magnetic moments, quarkonium decays,..., see e.g.
Ref.~\cite{Williams,DarkVector,Marciano12}), so the vector should be heavy, $m_{V}\gtrsim10$ GeV say, to evade or loosen them.

Actually, no matter its mass and the details of its fermionic couplings, we can immediately infer from the SM rates that the $h^{0}VZ$ and $h^{0}\gamma Z$ effective vertices arising from an SM fermion loop (or $W^{+}W^{-}$ loop, if $V$ couples to $W$) are too small to affect the $h^{0}$ total rate. Indeed, we can write $\Gamma(h^{0}\rightarrow\gamma V)\approx\Gamma(h^{0}\rightarrow ZV)\approx\bar{\lambda}^{2}\Gamma^{SM}(h^{0}\rightarrow\gamma Z)$ and $\Gamma(h^{0}\rightarrow VV)\approx\bar{\lambda}^{4}\Gamma^{SM}(h^{0}\rightarrow\gamma Z)$, up to phase-space corrections, since the $\gamma Z$ and $\gamma\gamma$ final states can be produced only through such loops. Given the SM rates in Table~\ref{Hwidth} and the bound $\bar{\lambda}<0.03$ from the $\rho$ parameter, these branching ratios are prohibitively small, $\mathcal{B}(h^{0}\rightarrow\gamma V)\approx\mathcal{B}(h^{0}\rightarrow ZV)\lesssim 10^{-6}$ and $\mathcal{B}(h^{0}\rightarrow VV)\lesssim10^{-9}$. The fermionic loops are thus subleading compared to the direct couplings to the Higgs boson, see Eq.~(\ref{RateGauge}).

\subsection{Gauge couplings}

The Higgs coupling scenario of the previous section is related to the kinetic
mixing scenario, defined by introducing the dark vector through the gauge
invariant coupling~\cite{Kinetic}:%
\end{subequations}
\begin{equation}
\mathcal{L}_{gauge}^{1}=\frac{\chi}{2}B_{\mu\nu}\times V^{\mu\nu
}\;.\label{KinMix}%
\end{equation}
Indeed, when the kinetic terms of the $B$ and $V$ fields are diagonalized
through the non-unitary transformation \cite{Kinetic,Foot} (we follow the
notations of Ref.~\cite{Williams}, to which we refer for more details)%
\begin{equation}
\left( \begin{array}[c]{c} B\\V \end{array} \right) \rightarrow
\left( \begin{array}[c]{cc} 1 & \sinh\eta\\ 0 & \cosh\eta \end{array} \right)  
\left( \begin{array}[c]{c} B\\V \end{array} \right) 
\;,\;\;\chi=\frac{\sinh\eta}{\cosh\eta}\;,
\end{equation}
one ends up with $V$ couplings to the SM particles aligned with the SM hypercharges. As explained at the end in the previous section, a unitary transformation then permits one to completely decouple the dark vector from the SM~\cite{Kinetic}.

When $\bar{m}_{V}>0$, this unitary transformation cannot be done. Instead, after putting the kinetic term in its canonical form, the mass matrix freezes the physical $V$ and $B$ states, and the $V$ field remains coupled to SM particles. To analyze those with the Higgs boson (which are missing in Ref.~\cite{Williams}) without going through the full diagonalization, let us assume $\chi\ll1$ and rewrite the kinetic mixing as%
\begin{equation}
\mathcal{L}_{gauge}^{1}=\chi c_{W}J_{\nu}^{em}\times V^{\nu}+\chi s_{W}Z_{\nu}\times\partial_{\mu}V^{\mu\nu}\;, 
\label{KinMix2}%
\end{equation}
where we integrated by parts, and set $\partial^{\mu}F_{\mu\nu}=-J_{\nu}^{em}$. The impact of the first term can be estimated as $\Gamma(h^{0}\rightarrow ZV)\approx\chi^{2}\Gamma(h^{0}\rightarrow Z\gamma)$, $\Gamma(h^{0}\rightarrow\gamma V)\approx\chi^{2}\Gamma(h^{0}\rightarrow\gamma\gamma)$, and $\Gamma(h^{0}\rightarrow VV)\approx\chi^{4}\Gamma(h^{0}\rightarrow\gamma\gamma)$, up to phase-space corrections. Because the electromagnetic field is not directly coupled to $h^{0}$, all these processes are very suppressed, even with $\chi\sim\mathcal{O}(1)$, and thus no constraints can be drawn from $\Gamma_{h}^{SM}$.

The second term appears more interesting since the $Z$ boson does have a tree-level coupling to the Higgs boson, see Eq.~(\ref{Zcoupl}). Treating the kinetic mixing as a mass insertion, the induced $V$ couplings to $h^{0}$ are, using $\partial_{\mu}V^{\mu\nu}=-m_{V}^{2}V^{\nu}$,%
\begin{equation}
\frac{M_{Z}^{2}H^{\dagger}H}{2v^{2}}Z_{\mu}Z^{\mu}\rightarrow\frac{M_{Z}^{2}H^{\dagger}H}{2v^{2}}\left(  Z_{\mu}Z^{\mu}+2\frac{\chi s_{W}r_{VZ}^{2}}{1-r_{VZ}^{2}}Z_{\mu}V^{\mu}+\left(  \frac{\chi s_{W}r_{VZ}^{2}}{1-r_{VZ}^{2}}\right)  ^{2}V_{\mu}V^{\mu}\right)  \;,
\end{equation}
with $r_{VZ}\equiv m_{V}/M_{Z}$ (the exact diagonalization of Ref.~\cite{Williams} leads to the same result when expanded around $\chi=0$). Once more, in the $m_{V}\rightarrow0$ limit, these couplings tend to zero, and so do the decay rates. They are obtained from those in Eq.~(\ref{RateGauge}) by substituting%
\begin{equation}
\bar{\lambda}\rightarrow\frac{\chi s_{W}}{1-r_{VZ}^{2}}\;.
\end{equation}
So, requiring $\Gamma_{h}$ to stay within 20\% of its SM value implies $\chi\lesssim0.7$ for $m_{V}<M_{h}-M_{Z}$. Since such large values are excluded by the $\rho$ parameter~\cite{Williams}, no visible impact on $\Gamma_{h}$ could arise. Despite of this, it should be remarked again that the $\Gamma(h^{0}\rightarrow ZV)$ could be our best window for this scenario since it is not invisible and could have a rate of the order of $\Gamma^{SM}(h^{0}\rightarrow \gamma\gamma)$, see Eq.~(\ref{RateGauge}).

\subsection{Effective couplings}

From the previous two scenarios, we can conclude that renormalizable and gauge-invariant couplings of the dark vector to the SM do not open new sizable Higgs boson decay channels, because a dark gauge invariance combined with the SM gauge invariance prevents a direct coupling to the Higgs boson. When the dark symmetry is softly broken by a $V$ mass term, as required to prevent this field from being rotated away, its direct couplings to the Higgs boson are proportional to the vector physical mass, and the decay rates are very small. Alternatively, the $V$ can couple to the Higgs boson at the loop level, but the fermion and $W^{+}W^{-}$ loops are very suppressed, and so are the rates. Hence, to get a visible impact on the total Higgs boson width, one needs a hard breaking of the dark gauge invariance, for example by decoupling the strength of the $H^{\dagger}H\times V_{\mu}V^{\mu}$ and $iH^{\dagger}\overleftrightarrow{\mathcal{D}}\hspace{0in}_{\mu}H\times V^{\mu}$ operators, as was analyzed in Secs.~2 and~3 (see also Ref.~\cite{Lebedev11} for an implementation based on the St\"{u}ckelberg mechanism).

In view of this, if the dark gauge invariance is unbroken or broken only
softly, then the largest effects could actually come from higher-dimensional
operators%
\begin{align}
\mathcal{H}_{eff-gauge}^{1}  &  =\frac{\eta_{1}}{\Lambda^{2}}H^{\dagger
}HB_{\mu\nu}\times V^{\mu\nu}+\frac{\tilde{\eta}_{1}}{\Lambda^{2}}H^{\dagger
}HB_{\mu\nu}\times\tilde{V}\hspace{0in}^{\mu\nu}\\
&  \;\;+\frac{\eta_{2}}{\Lambda^{2}}H^{\dagger}H\times V_{\mu\nu}V^{\mu\nu
}+\frac{\tilde{\eta}_{2}}{\Lambda^{2}}H^{\dagger}H\times V_{\mu\nu}\tilde
{V}\hspace{0in}^{\mu\nu}\\
&  \;\;+\frac{\eta_{3}}{\Lambda^{2}}H^{\dagger}\tau^{I}HW_{\mu\nu}^{I}\times
V^{\mu\nu}+\frac{\tilde{\eta}_{3}}{\Lambda^{2}}H^{\dagger}\tau^{I}HW_{\mu\nu
}^{I}\times\tilde{V}^{\mu\nu}+\frac{\eta_{4}}{\Lambda^{2}}iH^{\dagger
}\overleftrightarrow{\mathcal{D}}\hspace{0in}_{\nu}H\times\partial_{\mu}V^{\mu\nu}\;,
\end{align}
with $V_{\mu\nu}=\partial_{\mu}V_{\nu}-\partial_{\nu}V_{\mu}$, $\tilde
{V}\hspace{0in}^{\mu\nu}=\varepsilon^{\mu\nu\rho\sigma}V_{\rho\sigma}/2$.
These manifestly gauge-invariant interactions do not generate a vector mass,
but directly induce the $h^{0}\rightarrow VV$, $h^{0}\rightarrow ZV$, and
$h^{0}\rightarrow\gamma V$ processes. Thanks to the dark gauge invariance,
these rates are safe in the $m_{V}\rightarrow0$ limit%
\begin{subequations}
\begin{align}
\Gamma(h^{0}\overset{}{\rightarrow}VV)  &  =\frac{v^{2}}{8\pi M_{h}%
}\frac{M_{h}^{4}}{\Lambda^{4}}\frac{\beta_{V}}{4}(\eta_{2}^{2}(3+2\beta
_{V}^{2}+3\beta_{V}^{4})+8\tilde{\eta}_{2}^{2}\beta_{V}^{2})\;,\\
\Gamma(h^{0}\overset{}{\rightarrow}\gamma V)  &  =\frac{v^{2}}{8\pi M_{h}%
}\frac{M_{h}^{4}}{\Lambda^{4}}(1-r_{V}^{2})^{3}(\eta_{13}^{2}+\tilde{\eta
}_{13}^{2})\;,\\
\Gamma(h^{0}\overset{}{\rightarrow}ZV)  &  =\frac{v^{2}}{8\pi M_{h}%
}\frac{M_{h}^{4}}{\Lambda^{4}}\sqrt{\lambda}\left[  \eta_{31}^{2}(6r_{V}%
^{2}r_{Z}^{2}+\lambda)+\tilde{\eta}_{31}^{2}\lambda\frac{{}}{{}}\right. \\
&  \;\;\;\;\;\;\;\;\;\;\;\;\;\;\;\;\;\;\;\;\;\left.  +\frac{r_{V}^{2}}%
{2r_{Z}^{2}}(12r_{Z}^{2}(1-r_{V}^{2}-r_{Z}^{2})\bar{\eta}_{4}\eta
_{31}+(12r_{V}^{2}r_{Z}^{2}+\lambda)\bar{\eta}_{4}^{2})\right]  \;,
\end{align}
with $\lambda=\lambda(1,r_{Z}^{2},r_{V}^{2})$, $\bar{\eta}_{4}=g\eta
_{4}/2c_{W}$, $\eta_{13}=c_{W}\eta_{1}-s_{W}\eta_{3}$, $\eta_{31}=c_{W}%
\eta_{3}+s_{W}\eta_{1}$, and similarly for $\tilde{\eta}_{13,31}$. The scales
$\Lambda$ are then bounded as%
\end{subequations}
\begin{equation}%
\begin{array}
[c]{cll}%
\Gamma(h^{0}\rightarrow VV)<\Gamma_{h}^{SM}/5\;\;\Rightarrow & \Lambda
\gtrsim1.9\;\text{TeV} & \;(\eta_{2}\text{, }\tilde{\eta}_{2}\sim
\mathcal{O}(1))\;,\\
\Gamma(h^{0}\rightarrow\gamma V)<\Gamma_{h}^{SM}/5\;\;\Rightarrow &
\Lambda\gtrsim1.5\;/\;1.1\;\text{TeV} & \;(\eta_{1}\text{, }\tilde{\eta}%
_{1}\;/\;\eta_{3}\text{, }\tilde{\eta}_{3}\sim\mathcal{O}(1))\;,\\
\Gamma(h^{0}\rightarrow ZV)<\Gamma_{h}^{SM}/5\;\;\Rightarrow & \Lambda
\gtrsim0.6\;/\;0.8\;/\;0.2\;\text{TeV} & \;(\eta_{1}\text{, }\tilde{\eta}%
_{1}\;/\;\eta_{3}\text{, }\tilde{\eta}_{3}\;/\;\eta_{4}\sim\mathcal{O}(1))\;.
\end{array}
\label{boundGI}%
\end{equation}
The vector is taken as massless for all the bounds except $\eta_{4}$. Since
$\partial_{\mu}V^{\mu\nu}$ vanishes for $m_{V}=0$, it is given for
$m_{V}=25\;$GeV. Actually, the $\eta_{4}$ operator can be matched onto the
$Z_{\mu}V^{\mu}$ couplings studied in the previous two scenarios, see e.g. the
last coupling of Eq.~(\ref{KinMix2}), so it is expected to remain
unconstrained by the total SM Higgs decay width (remember that consistency
requires $\Lambda>v\approx246$ GeV). Also, it must be remarked that the first
operator regenerates the kinetic mixing~(\ref{KinMix}), with $\chi=\eta
_{1}v^{2}/\Lambda^{2}\lesssim0.03$. Low-energy constraints on $\chi$, and
thereby on $\eta_{1}v^{2}/\Lambda^{2}$, are thus much stronger when $m_{V}$ is
light (see e.g. Ref.~\cite{Williams}). For larger masses, $m_{V}>10$~GeV say,
these other bounds are either evaded or satisfied for $\chi\lesssim0.03$. The
other couplings are not constrained yet, because the $\eta_{2}v^{2}%
/\Lambda^{2}$ term induces an innocuous wavefunction renormalization, while
those proportional to $\tilde{\eta}_{1,2}v^{2}/\Lambda^{2}$ are total
derivatives and can be dropped~\cite{U1Mono}.

This effective operator approach can be extended to spin~$3/2$ states, for
which it is impossible to construct renormalizable gauge invariant couplings
to the SM. For instance, the leading manifestly gauge-invariant operators
involving $H^{\dagger}H$ are%
\begin{subequations}
\begin{equation}
\mathcal{H}_{eff-gauge}^{3/2}=\frac{c_{S}^{\prime}}{\Lambda^{3}}H^{\dagger
}H\times\overline{\Psi}\hspace{0in}^{\mu\nu}\Psi_{\mu\nu}+\frac{c_{P}^{\prime
}}{\Lambda^{3}}H^{\dagger}H\times\overline{\Psi}\hspace{0in}^{\mu\nu}%
\gamma_{5}\Psi_{\mu\nu}+...\;,
\end{equation}
with $\Psi_{\mu\nu}\equiv\partial_{\mu}\Psi_{\nu}-\partial_{\nu}\Psi_{\mu}$.
As for the dark vector case, these couplings do not correct $m_{\Psi}$ and the
decay rates are automatically safe in the $m_{\Psi}\rightarrow0$ limit thanks
to the derivatives occurring in $\Psi_{\mu\nu}$
\end{subequations}
\begin{equation}
\Gamma(h^{0}\rightarrow\Psi\Psi)=\frac{v^{2}\beta_{\Psi}}{8\pi M_{h}}%
(c_{S}^{\prime2}\beta_{\Psi}^{2}\bar{\beta}_{\Psi}^{\prime}+c_{P}^{\prime
2}\bar{\beta}_{\Psi}^{\prime\prime})\frac{10M_{h}^{6}}{9\Lambda^{6}}\;,
\end{equation}
with $\bar{\beta}_{\Psi}^{\prime}=(5+6\beta_{\Psi}^{2}+9\beta_{\Psi}^{4})/20$
and $\bar{\beta}_{\Psi}^{\prime\prime}=(9+6\beta_{\Psi}^{2}+5\beta_{\Psi}%
^{4})/20$. Setting $m_{\Psi}=0$, the tiny SM Higgs boson width implies%
\begin{equation}
\Gamma(h^{0}\rightarrow\Psi\Psi)<\Gamma_{h}^{SM}/5\;\;\Rightarrow
\Lambda\gtrsim0.7\;\text{TeV}\;.
\end{equation}
Though this scale is rather low, it must be stressed that rare decays
typically produce looser bounds because gauge-invariant FCNC operators are at
least of dimension eight~\cite{KamenikS11}. So, $\Gamma_{h}$ may actually be
our best window in case $\Psi$ is required to couple in a manifestly
gauge-invariant way.

\section{Fermionic channels}

Let us now turn to the decays involving SM fermions together with dark states,
starting with those which conserve lepton and baryon numbers.

\subsection{Baryon and lepton number conserving channels}

Because of the required $SU(3)_{C}\otimes SU(2)_{L}\otimes U(1)_{Y}$ invariance,
there are two kinds of fermionic operators. The couplings of the dark states
to scalar or tensor quark and lepton currents always involve a Higgs
field,
\begin{equation}
\Gamma^{S}=H^{\dagger}\bar{D}Q,\;\;H^{\dagger}\bar{E}L,\;\;H^{\ast\dagger
}\bar{U}Q,\;\;\;\;\Gamma_{\mu\nu}^{T}=H^{\dagger}\bar{D}\sigma_{\mu\nu
}Q,\;\;H^{\dagger}\bar{E}\sigma_{\mu\nu}L,\;\;H^{\ast\dagger}\bar{U}\sigma
_{\mu\nu}Q\;,
\end{equation}
while those to the vector and axial vector currents need an additional $H^{\dagger}H$ pair to contribute to $h^{0}$ decays,%
\begin{equation}
\Gamma_{\mu}^{V}=H^{\dagger}H\bar{Q}\gamma_{\mu}Q,\;\;H^{\dagger}H\bar
{D}\gamma_{\mu}D\;,\;\;H^{\dagger}H\bar{L}\gamma_{\mu}L\;,\;\;H^{\dagger
}H\bar{E}\gamma_{\mu}E\;,
\end{equation}
where the triplet $SU(2)_{L}$ contractions for fermionic doublets are
implicitly included. In terms of these currents, the simplest operators are
(the hermitian conjugate for the operators with $\Gamma^{S}$ and $\Gamma^{T}$
are understood)
\begin{subequations}
\label{HFermion}%
\begin{align}
\mathcal{H}_{eff}^{0} &  =\frac{g_{S}^{\phi}}{\Lambda}\Gamma^{S}\times
\phi+\frac{g_{S}^{\phi\phi}}{\Lambda^{2}}\Gamma^{S}\times\phi^{\dagger}%
\phi+\frac{g_{V}^{\phi}}{\Lambda^{3}}\Gamma_{\mu}^{V}\times\partial^{\mu}%
\phi+\frac{ig_{V}^{\phi\phi}}{\Lambda^{4}}\Gamma_{\mu}^{V}\times\phi^{\dagger
}\overleftrightarrow{\partial}\hspace{0in}^{\mu}\phi\;,\\
\mathcal{H}_{eff}^{1/2} &  =\frac{f_{L,R}^{S}}{\Lambda^{3}}\Gamma^{S}%
\times\bar{\psi}_{L,R}\psi_{R,L}+\frac{f^{T}}{\Lambda^{3}}\Gamma_{\mu\nu}%
^{T}\times\bar{\psi}_{R}\sigma^{\mu\nu}\psi_{L}+\frac{f_{L,R}^{V}}{\Lambda
^{4}}\Gamma_{\mu}^{V}\times\bar{\psi}_{L,R}\gamma^{\mu}\psi_{L,R}\;,\\
\mathcal{H}_{eff}^{1} &  =\frac{h^{T}}{\Lambda^{2}}\Gamma_{\mu\nu}^{T}\times
V^{\mu\nu}+\frac{h^{V}}{\Lambda^{2}}\Gamma_{\mu}^{V}\times V^{\mu}\;,\\
\mathcal{H}_{eff}^{3/2} &  =\frac{f_{S,P}^{S}}{\Lambda^{3}}\Gamma^{S}%
\times\overline{\Psi}\hspace{0in}^{\mu}(1,\gamma_{5})\Psi_{\mu}+\frac{f_{S,P}%
^{T}}{\Lambda^{3}}\Gamma_{\mu\nu}^{T}\times\overline{\Psi}\hspace{0in}^{[\mu
}(1,\gamma_{5})\Psi^{\nu]}+\frac{f_{T}^{T}}{\Lambda^{3}}\Gamma_{\mu\nu}%
^{T}\times\overline{\Psi}\hspace{0in}_{\rho}\sigma^{\mu\nu}\Psi^{\rho
}\nonumber\\
&  \;\;\;\;+\frac{f_{V,A}^{V}}{\Lambda^{4}}\Gamma_{\mu}^{V}\times
\overline{\Psi}\hspace{0in}^{\rho}\gamma^{\mu}(1,\gamma_{5})\Psi_{\rho}\;\;.
\end{align}

Most of these interactions do not appear very promising for several reasons.
Firstly, they have high dimensions compared to those studied in the previous
sections, and they involve many particles so the decay rates are significantly
phase-space suppressed. This is apparent in Table~\ref{HBRff}, with only the
$h^{0}\rightarrow f\bar{f}\phi$ and $h^{0}\rightarrow f\bar{f}V$ channels,
induced by $g_{S}^{\phi}$ and $h^{T}$, potentially large enough. Note that the
magnetic operator tuned by $h^{T}$ is gauge invariant, so the limit
$m_{V}\rightarrow0$ is safe. Also, remember that the $g_{S}^{\phi}$ coupling
is forbidden if the scalar is charged, or if a $\mathbb{Z}_{2}$ symmetry
$\phi\rightarrow-\phi$ is enforced, as done to remove the potentially much
larger effects from the $\mu^{\prime}$ coupling of Eq.~(\ref{HHss}).%

\begin{table}[t] \centering
$
\begin{tabular}
[c]{|ll|ll|ll|ll|}\hline
\multicolumn{2}{|l|}{Spin 0} & \multicolumn{2}{|l|}{Spin 1/2} &
\multicolumn{2}{|l|}{Spin 1} & \multicolumn{2}{|l|}{Spin 3/2}\\\hline
$g_{S}^{\phi}$ & $0.062$ & $f_{L,R}^{S}$ & $3.7\cdot10^{-8}$ & $h^{T}$ &
$0.0062$ & $f_{S,P}^{S}$ & $10^{-10}(^{\ast\ast})$\\
$g_{V}^{\phi}$ & $4\cdot10^{-4}$ & $f_{L,R}^{V}$ & $5.4\cdot10^{-8}$ & $h^{V}$
& $0.0015(^{\ast})$ & $f_{S,P}^{T}$ & $10^{-10}(^{\ast\ast})$\\
$g_{S}^{\phi\phi}$ & $2.6\cdot10^{-6}$ & $f^{T}$ & $6.0\cdot10^{-7}$ &  &  &
$f_{T}^{T}$ & $10^{-9}(^{\ast\ast})$\\
$g_{V}^{\phi\phi}$ & $1.4\cdot10^{-8}$ &  &  &  &  & $f_{V,A}^{V}$ &
$10^{-10}(^{\ast\ast})$\\\hline
\end{tabular}
\ \ \ $
\caption{Branching ratios for $h^{0}\rightarrow ff+\slashed E$, as induced by the operators in Eq.~(\ref{HFermion}). We set $\Lambda=500$ GeV, $M_{h}=125$ GeV, $\Gamma_{h}=4$ MeV, the masses of the SM fermions and of the dark particles to zero, and each Wilson coefficient to one in turn. Summation over the $N_{c}=3$ colors for quark states, as well as over fermion species, has not been done. If all final states are produced equally, even when flavor-violating, these branching ratios should be multiplied by $6N_{c}+3N_{c}+6=33$. In that case, the width for the modes induced by $g_{S}^{\phi}$ and $h^{T}$ become significant compared to $\Gamma_{h}^{SM}$. For the non-gauge invariant operators, we use Eq.~(\ref{VVPP}) and set $(^{\ast})$
$m_{V}=h^{V}v$ and $(^{\ast\ast})$ $\Lambda=v^{2}/(2m_{\Psi})$. In this latter case, we quote the branching ratios for $m_{\Psi}=35$ GeV, where they are maximum. Note that for both the dark spin 1 and 3/2 states, these electroweak masses lead to
very suppressed decay rates (compare with Eq.~(\ref{HHVPbnd})) because of the high dimensionality of the operators.}
\label{HBRff}
\end{table}

Secondly, we cannot expect a precise measurement of all the fermionic decay channels in the near future. In addition, most of these operators are obscured either by an SM process, or by the processes induced by the lower-dimensional operators considered in the previous sections. To illustrate this, consider first the operators involving $\Gamma_{\mu}^{V}$. When flavor-diagonal, they are directly obtained from those in Eq.~(\ref{GaugeOp}) by coupling the $Z$ boson to SM fermions (see Fig.~\ref{SMvsNP}$a$). Those processes are suppressed by $1/M_{Z}^{2}$ when the $Z$ is off-shell, to be compared with the
additional $1/\Lambda^{2}$ power introduced in Eq.~(\ref{HFermion}). Further, a reasoning similar to that done following Eq.~(\ref{SMampli}) shows that all the vector operators producing pairs of dark particles are not competitive compared to $h^{0}\rightarrow Z^{\ast}Z^{\ast}\rightarrow f\bar{f}\nu\bar{\nu}$. They have the same experimental signatures, but $\Gamma(h^{0}\rightarrow ff^{\prime}+\slashed E)<\Gamma^{SM}(h^{0}\rightarrow Z^{\ast}Z^{\ast})$ for $\Lambda>v$ (the same applies to the $h^{0}\rightarrow W^{\ast}W^{\ast}$ processes for leptonic final states, see Fig.~\ref{SMvsNP}$b$).

The situation is similar for the scalar and most tensor operators, though the
flavor structure of the Wilson coefficients here plays a role. If we assume
that the Minimal Flavor Violation (MFV) ansatz~\cite{MFV} is valid, all the
flavor violation is induced by the SM dynamics. This permits one to evade the
tight constraints derived from rare $K$ and $B$ decays when the dark particle
is light~\cite{KamenikS11}. But then, the chirality flip is tuned by the SM
fermion masses, for example $\bar{D}^{I}\mathbf{Y}_{d}Q^{J}\rightarrow
\bar{d}_{R}^{I}\mathbf{m}_{d}d_{L}^{J}/v$, with $\mathbf{m}_{d}$ the diagonal
down quark mass matrix. These are precisely the SM couplings of the Higgs
boson to SM fermions. So, all the scalar and some of the tensor effective
operators of Eq.~(\ref{HFermion}) appear as $\mathcal{O}(M_{h}^{2}/\Lambda
^{2})$ corrections to the operators of Sec.~2, as induced by the $h^{0}h^{0}$
term of Eq.~(\ref{SSBHH}), see Fig.~\ref{SMvsNP}$c$. In this respect, the
magnetic $h^{T}\Gamma_{\mu\nu}^{T}\times V^{\mu\nu}$ operator appears again as
the most promising, because it is entirely independent from previously
considered operators.

In summary, the only accessible operators are those inducing the
$h^{0}\rightarrow f\bar{f}^{\prime}\phi$ and $h^{0}\rightarrow f\bar
{f}^{\prime}V$ modes, with $f\bar{f}^{\prime}=d^{I}\bar{d}^{J},\ell^{I}%
\bar{\ell}^{J},u^{K}\bar{u}^{L},$ $I,J=1,2,3$, $K,L=1,2$ the flavor indices.
If the scale $\Lambda$ is at or below the TeV scale and if the Wilson coefficients
are generic, so that the required chirality flip is not induced by the fermion
masses, then they are large enough to show up in the $h^{0}\rightarrow
Z^{\ast}Z^{\ast}\rightarrow f\bar{f}\nu\bar{\nu}$ decay channels and, for
leptonic final states, in the $h^{0}\rightarrow W^{\ast}W^{\ast}%
\rightarrow\ell\bar{\nu}_{\ell}\bar{\ell}^{\prime}\nu_{\ell^{\prime}}$ decay channels.

\subsection{Baryon and lepton number violating channels}

Assuming the dark particles are colorless, the operators violating baryon
number ($\mathcal{B}$) require at least three quark fields~\cite{KamenikS11},
and thus have too high dimensions to play any role in $h^{0}$ decays. By
contrast, those violating lepton number ($\mathcal{L}$) are constructed out of
the simple field combination%
\end{subequations}
\begin{equation}
HL\rightarrow\frac{1}{\sqrt{2}}\left(  v+h^{0}\right)  \nu_{\ell}\;,\label{HL}%
\end{equation}
which directly couples the Higgs boson to neutrinos. The simplest such
operators are~\cite{KamenikS11}
\begin{subequations}
\label{HeffDL}%
\begin{align}
\mathcal{H}_{eff}^{0} &  =\frac{a_{1}}{\Lambda^{2}}H\bar{L}^{C}LH\times
\phi+\frac{a_{2}}{\Lambda^{3}}H\bar{L}^{C}LH\times\phi^{\dagger}\phi+h.c.\;,\\
\mathcal{H}_{eff}^{1/2} &  = b_{0}H\times\bar{\psi}_{R}L+\frac{b_{1}}%
{\Lambda^{2}}B_{\mu\nu}H\times\bar{\psi}_{R}\sigma^{\mu\nu}L+\frac{b_{2}%
}{\Lambda^{2}}W_{\mu\nu}^{I}H\tau^{I}\times\bar{\psi}_{R}\sigma^{\mu\nu
}L+h.c.\;,\\
\mathcal{H}_{eff}^{1} &  =\frac{c_{1}}{\Lambda^{3}}H\bar{L}^{C}\mathcal{D}%
_{\mu}LH\times V^{\mu}+\frac{c_{2}}{\Lambda^{3}}H\bar{L}^{C}L\mathcal{D}_{\mu
}H\times V^{\mu}+\frac{c_{3}}{\Lambda^{3}}H\bar{L}^{C}\sigma_{\mu\nu}LH\times
V^{\mu\nu}+h.c.\;,\\
\mathcal{H}_{eff}^{3/2} &  =\frac{d_{0}}{\Lambda}\mathcal{D}_{\mu}H\times\overline{\Psi
}\hspace{0in}^{\mu}L+\frac{d_{1}}{\Lambda^{2}}B_{\mu\nu}H\times\overline{\Psi
}\hspace{0in}^{[\mu}\gamma^{\nu]}L+\frac{d_{2}}{\Lambda^{2}}W_{\mu\nu}%
^{I}H\tau^{I}\times\overline{\Psi}\hspace{0in}^{[\mu}\gamma^{\nu]}L+h.c.\;.
\end{align}
Because the neutrinos are fermions, the $\Delta\mathcal{L}$ operators for dark
scalar or vector states must involve twice the combination~(\ref{HL}). They
are strongly suppressed by their higher dimensionality and by the three or
four body phase-space, hence need not be considered anymore. On the other
hand, the leading operators for dark fermions can have low dimensions. In the
spin~1/2 case, this embodies the so-called neutrino portal~\cite{NeutPortals}.
Let us note also that strictly speaking, it suffices to assign a non-zero
$\mathcal{L}$ to the dark fermions for those operators to be $\Delta
\mathcal{L}=0$. In this sense, it is not sufficient to enforce the SM global
symmetries to discard them.

For the dark spin~1/2 state, the leading renormalizable operator cannot have any impact on $\Gamma_{h}$, because $b_{0}$ must be tiny to avoid inducing a too large neutrino mass\footnote{This holds even if $\psi$ is a heavy fourth generation neutrino, as in Ref.~\cite{4Gneut}. Indeed, by assumption, the $L$ fields in Eq.~(\ref{HeffDL}) stand for SM fields, i.e., only the first three generations. So $b_0$ is necessarily bounded by the light neutrino masses. This also means that the $h^0\rightarrow\nu_4\nu_4$ process is not induced by the $b_0$ operator. Rather, it could originate from the four-generation $H\bar{\nu}_{R,4}L_{4}$ (Dirac) or $H\bar{L}_{4}^{C}L_{4}H$ (Majorana) mass operator. The correlation between the electroweak mass correction and the Higgs boson coupling is then the same as that studied in Sec.~2.1, see Eq.~(\ref{sspp}).}. Let us thus consider the non-renormalizable $W_{\mu\nu}^{I}H\tau^{I}\times\bar{\psi}_{R}\sigma^{\mu\nu}L$ and $B_{\mu\nu}H\times\bar{\psi}_{R}\sigma^{\mu\nu}L$ operators, which drive $h^{0}\rightarrow W^{+}\ell^{-}\psi$, $h^{0}\rightarrow Z\nu\psi$, and $h^{0}\rightarrow\gamma\nu\psi$. The first two processes do not offer interesting windows because they cannot be distinguished from the $h^{0}\rightarrow W^{+}W^{-}[\rightarrow\ell^{-}\bar{\nu}]$ and $h^{0}\rightarrow ZZ[\rightarrow\nu\bar{\nu}]$ transitions. A reasoning similar to that following Eq.~(\ref{SMampli}) shows that $\Gamma(h^{0}\rightarrow W^{+}\ell^{-}\psi)<\Gamma^{SM}(h^{0}\rightarrow W^{+}\ell^{-}\bar{\nu})$ and $\Gamma(h^{0}\rightarrow Z\nu\psi)<\Gamma^{SM}(h^{0}\rightarrow Z\nu\bar{\nu})$ when $\Lambda>v$, see Figs.~\ref{SMvsNP}$a$ and~\ref{SMvsNP}$b$.

The situation for $h^{0}\rightarrow\gamma\nu\psi$ is different since there is
no tree-level $h^{0}Z\gamma$ vertex, and the $h^{0}\rightarrow\gamma
Z[\rightarrow\nu\bar{\nu}]$ rate is very suppressed in the SM, see
Table~\ref{Hwidth}. On the other hand, for $m_{\psi}=0$ (thus discarding the
operator of Eq.~(\ref{HHpp})), summing over the three neutrino flavors, and assuming lepton universality for the $b_{1,2}$ coefficients:%
\end{subequations}
\begin{equation}
\mathcal{B}(h^{0}\rightarrow\gamma\nu\psi)=3\tau_{h}\frac{M_{h}^{5}(c_{W}b_{1}-s_{W}b_{2})^{2}}{640\pi^{3}\Lambda^{4}}\;\;\overset{b_{i}\sim \mathcal{O}(1)}{=}(14\;,\;2\;,\;0.1)\,\%\;\;\text{for }\Lambda=(0.3\;,\;0.5\;,\;1)\,\text{TeV\ },
\end{equation}
which is larger than $\mathcal{B}(h^{0}\rightarrow\gamma Z)$ or $\mathcal{B}(h^{0}\rightarrow\gamma\gamma)$ when $\Lambda\lesssim850$ GeV, but remain smaller than $\mathcal{B}(h^{0}\rightarrow WW^{\ast},ZZ^{\ast})$ for all values $\Lambda>v$, in good agreement with the expected $\mathcal{O}(v^{2}/\Lambda^{2})$ suppression of the amplitude mentioned above. Whether
such a signal can be seen would require a detailed analysis. For now, we just conclude that the total SM Higgs boson width does not constrain the spin~1/2 operators in Eq.~(\ref{HeffDL}).

\begin{figure}[t]
\centering                           \includegraphics[width=7cm]{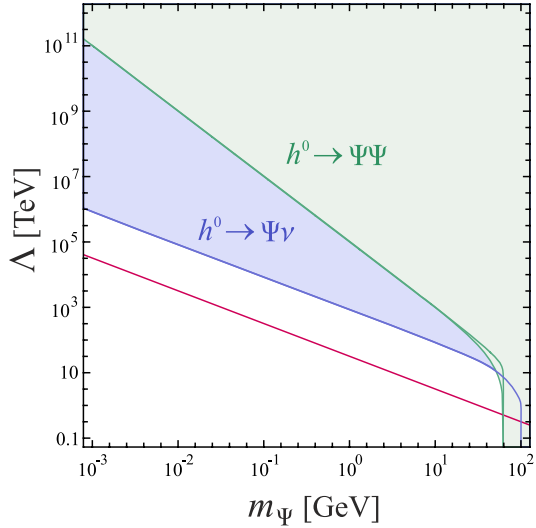}
\caption{Region in the $\Lambda-m_{\Psi}$ plane allowed by requiring $\Gamma(h^{0}\rightarrow\Psi\nu)<\Gamma_{h}^{SM}/5$ [blue], as induced by the operator of Eq.~(\ref{HeffDL}). We have set $d_{0}^{\ell}=1$ and summed over $\ell= e, \mu, \tau$ and $\Psi\nu\equiv\Psi\bar{\nu}+\overline{\Psi}\nu$. The allowed area for the $h^{0}\rightarrow\Psi\Psi$ mode [green], as induced by the operator of Eq.~(\ref{HHPP}) and shown in Fig.~\ref{FigHHVPsi}, is here plotted using the logarithmic mass scale. The red line denotes $m_{\Psi}=v^{2}/(2\Lambda)$, and lies entirely out of the allowed regions, except for masses just below or anywhere above the kinematical endpoints.}%
\label{FigHHPsiNu}%
\end{figure}

For the dark spin~3/2 state, the next-to-leading operators lead to the same signatures as for dark spin~$1/2$ states. However, the leading operator $\mathcal{D}_{\mu}H\times\overline{\Psi}\hspace{0in}^{\mu}L$ is no longer constrained by neutrino masses, and induces the invisible $h^{0}\rightarrow\Psi\nu$ decay. The rate is, for each neutrino flavor $\ell= e, \mu, \tau$:%
\begin{equation}
\Gamma(h^{0}\rightarrow\Psi\nu_{\ell})=\frac{M_{h}^{3}}{96\pi\Lambda^{2}}\frac{(1-r_{\Psi}^{2})^{4}}{r_{\Psi}^{2}}|d_{0}^{\ell}|^{2}\;.
\end{equation}
Inspired by Eq.~(\ref{VVPP}), the $m_{\Psi}^{-2}$ singularity is cured by setting $\Lambda=v^{2}/(2m_{\Psi})$, so that%
\begin{equation}
\Gamma(h^{0}\rightarrow\Psi\nu_{\ell})=\frac{M_{h}^{5}}{24\pi v^{4}}(1-r_{\Psi}^{2})^{4}|d_{0}^{\ell}|^{2}\approx(110\;\text{MeV})\times(1-r_{\Psi}^{2})^{4}|d_{0}^{\ell}|^{2}\;.
\end{equation}
This is too large compared to $\Gamma_{h}^{SM}$ for all values of the mass except
for a tiny range close to the kinematical threshold, now at $m_{\Psi}=M_{h}$.
This shows that the tiny $\Gamma_{h}^{SM}$ is incompatible with a dark
spin~3/2 particle coupled through the neutrino portal, at least as long as its
mass is generated by electroweak scale physics as $m_{\Psi}=v^{2}/2\Lambda$.

To generalize this result to arbitrary masses, we follow the same strategy as in the previous sections and introduce a Lagrangian mass term $\bar{m}_{\Psi}$. Then, the exclusion region in the $\Lambda-m_{\Psi}$ plane is shown in Fig.~\ref{FigHHPsiNu}. Though the Higgs portal operator $H^{\dagger}H\overline{\Psi}\hspace{0in}^{\mu}\Psi_{\mu}$ leads to even tighter constraints on $\Lambda$, the bounds on the scale are comparable for $m_{\Psi}\gtrsim5$ GeV. In any case, light spin~3/2 states are clearly incompatible with small NP scales, at least when coupled to the SM through these effective interactions.

\section{Summary and conclusion}

In this paper, the decays of the Higgs boson to new light neutral particles of
spin~0, $1/2$, $1$, and $3/2$, respectively denoted $\phi$, $\psi$, $V$, and
$\Psi$, were systematically analyzed. We have included all the leading
effective operators, whether renormalizable or not, and characterized their
possible signatures. These dark particles were not assumed stable but only
sufficiently long-lived to escape as missing energy. They are thus not
necessarily viable dark matter candidates, and the corresponding constraints
from direct detection or cosmology were not imposed. On the other hand, we
have shown that the tiny SM width $\Gamma_{h}^{SM}$ of a light Higgs boson
already suffices to derive strong constraints on their couplings to the SM.

\begin{figure}[t]
\centering         \includegraphics[width=16cm]{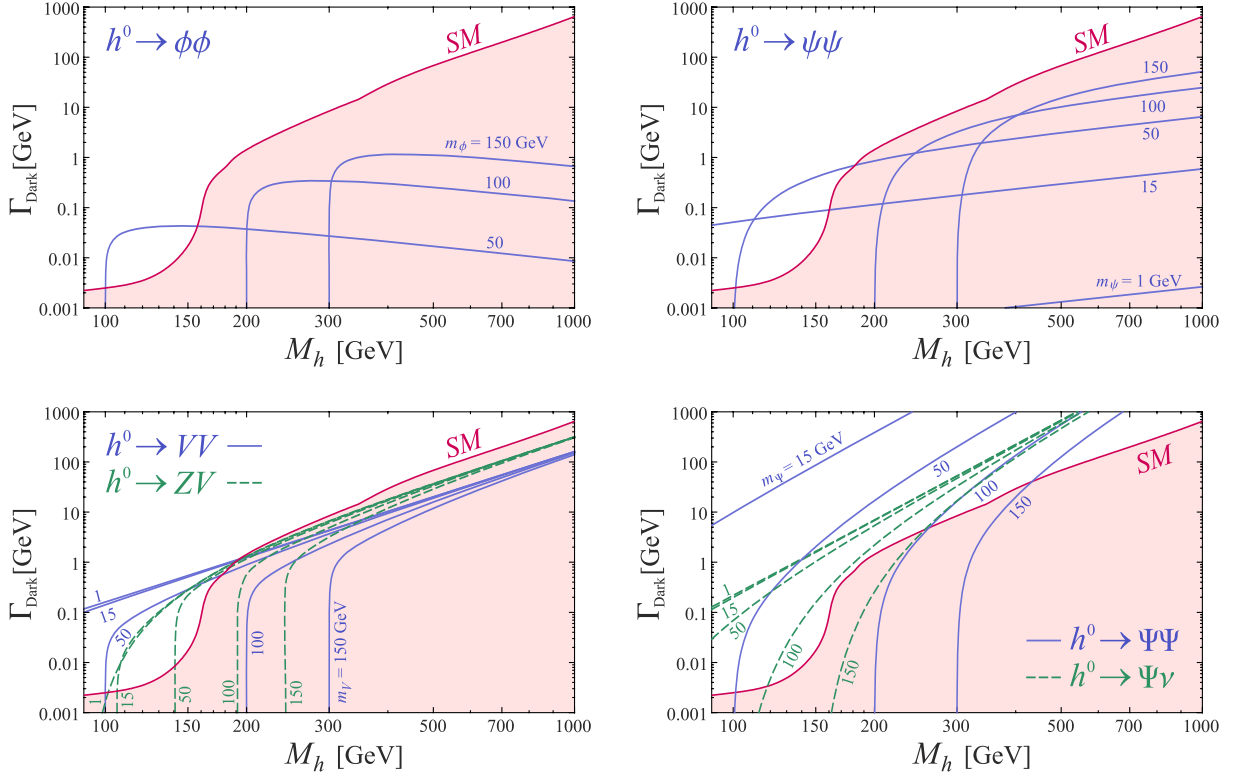}
\caption{Comparison between the dominant rates into dark particles and the total SM Higgs boson width, as a function of the Higgs boson mass $M_{h}$ and setting the Wilson coefficients of the relevant operators to one. The dark rates are plotted assuming purely electroweak dark particle masses, and thus represents natural upper bounds. Indeed, if one tries to increase a given dark rate by enhancing the coupling to the Higgs boson, then the dark particle gets more massive and the curve also shifts to the right. The only way to prevent this would be to allow for a strong cancellation between the dark and electroweak contributions to the dark particle mass, i.e., to fine-tune them. So, barring this, dark scalars or fermions can never hide a heavy Higgs boson, while dark vectors and spin 3/2 states could, but only when the breaking of the dark gauge invariance is hard.}%
\label{FigHtotal}%
\end{figure}

Thanks to their mild phase-space suppression, the two-body invisible
$h^{0}\rightarrow\phi\phi$, $\psi\psi$, $VV$, $\Psi\Psi$ decays offer the best
windows (see Fig.~\ref{FigHtotal}). Further, their rates are tightly
correlated with the electroweak contributions to the masses of these
particles, a fact we have used to set limits on the physical masses of these
states under various scenarios. We have also identified other two-body decays
of interest. Firstly, the $h^{0}\rightarrow\Psi\nu$ decay could actually be
our prime window for dark spin~3/2 states (see Fig.~\ref{FigHtotal}),
especially if $\Psi$ carries a non-zero lepton number. Secondly, the
$h^{0}\rightarrow ZV$ decay is a competitive, partially visible channel to
search for dark vectors. It would show up as a monochromatic peak over the
three-body SM $h^{0}\rightarrow ZZ^{\ast}[\rightarrow\nu\bar{\nu}]$ process.
Even though a precise measurement of the latter is challenging, a future
experimental bound on $\Gamma(h^{0}\rightarrow Z+\slashed E)$ could be far tighter than on $\Gamma(h^{0}\rightarrow \slashed E)$. Finally, the $h^{0}\rightarrow\gamma V$ decay mode does not appear
competitive, because it proceeds either at the loop level, or through higher
dimensional operators.

It should be stressed that for dark spin~1 and spin~3/2 particles, the fate of
the dark gauge symmetry determines the strength of the possible signals. For
dark vectors, we identified three different setups. First, if this symmetry is
unbroken, the only couplings to $h^{0}$ originate from non-renormalizable
operators. A 20\% enhancement over $\Gamma_{h}^{SM}$ requires a rather low NP
scale, at around $1$ TeV. Second, if this symmetry is broken only softly by
the dark vector mass, the renormalizable $h^{0}VV$ and $h^{0}VZ$ couplings end
up proportional to this soft breaking term, and the decay rates are too small
to affect $\Gamma_{h}^{SM}$ (though $\mathcal{B}(h^{0}\rightarrow ZV)$ could
reach a few per mil). Third, it is only in the presence of a hard breaking of
the dark gauge invariance that $\Gamma(h^{0}\rightarrow VV)$ and/or
$\Gamma(h^{0}\rightarrow ZV)$ can compete with, and even surpass, the SM Higgs
boson decay rates. For dark spin~3/2 particles, the situation is simpler
because it is not possible to construct renormalizable couplings of $\Psi$
with the SM. A strict enforcement of the dark gauge symmetry on the effective
operators prevents any effect for a NP scale above about $1$ TeV, while
$\Gamma(h^{0}\rightarrow\Psi\Psi)$ and/or $\Gamma(h^{0}\rightarrow\Psi\nu)$
can be huge even for extremely high NP scales when the effective operators
explicitly break this symmetry (see Fig.~\ref{FigHHPsiNu}).

By contrast, we found most processes with SM fermions in the final states to
be too small to be observed. This stems from their experimental entanglement
with the $h^{0}\rightarrow Z^{(\ast)}Z^{\ast}\ $and $h^{0}\rightarrow
W^{(\ast)}W^{\ast}$ decays, see Fig.~\ref{SMvsNP}. With the NP scale above the
electroweak scale, the SM rates are always larger. The only exceptions,
besides $h^{0}\rightarrow\Psi\nu$, are the three body decays $h^{0}\rightarrow
f\bar{f}^{\prime}\phi$, $h^{0}\rightarrow f\bar{f}^{\prime}V$, $h^{0}%
\rightarrow\gamma\nu\psi$, and $h^{0}\rightarrow\gamma\nu\Psi$, thanks to the
milder phase-space suppression and lower dimension of the effective operators.
Still, none of these modes could be dominant, so they would require dedicated
searches to be competitive with two-body processes.

To close our analysis, let us go back and relax our starting assumption.
Specifically, throughout the paper, the recent hint~\cite{Hints} of a Higgs
boson at around $M_{h}=125$ GeV, mainly in the $\gamma\gamma$ channel, was
used to set a bound on the width of the Higgs boson. Indeed, such a signal
should not have been seen if the total Higgs boson width were enhanced by some
new decay channels. But, if this signal is not confirmed, the exclusion range
for the SM Higgs boson mass would essentially extend all the way up to about
$600$ GeV.

In this perspective, the presence of a light dark state coupled to $h^{0}$
could play a crucial role. Clearly, a new large decay channel would invalidate
the exclusion range, since it would significantly suppress the branching
ratios to SM final states. To check if and when this is possible, we plotted
in Fig.~\ref{FigHtotal} the dominant $h^{0}$ decay rates to dark states as a
function of the Higgs boson mass. As can be seen, the rates get larger as the
dark particle spin increases, whether the leading operators are renormalizable
($\phi$ and $V$) or not ($\psi$ and $\Psi$). Note that the plotted dark rates
can be understood a upper bounds, assuming the absence of a strong fine-tuning
between the dark sector and electroweak contributions to the dark particle
masses (the latter originating from Eq.~(\ref{HH})).

While any type of dark particles could hide a light Higgs boson, a fact
extensively used in this paper, this is not true for larger masses.
Specifically, the $h^{0}\rightarrow\phi\phi$ and $h^{0}\rightarrow\psi\psi$
rates are always too small compared to the SM width for $M_{h}\gtrsim160$ GeV
and $M_{h}\gtrsim180$ GeV, respectively. For heavier Higgs boson masses, only
dark vector and spin~3/2 particles could have hampered the Higgs boson
searches through SM decay channels, and this provided first that their
couplings to the Higgs boson explicitly break a dark gauge invariance, and
second that their physical masses are of the order of the electroweak
contributions stemming from Eq.~(\ref{HH}). This clearly shows that the tiny
SM width of a light Higgs boson could be a gift from Nature, allowing us to
probe for the presence of yet unknown relatively light neutral particles with
an unprecedented sensitivity.

\subsection*{Acknowledgments}

C.S. would like to thank S. Davidson and P. Verdier for interesting
discussions. This work is supported in part by the Slovenian Research Agency.

\end{document}